\documentclass[aps,twocolumn,amsmath,amssymb,longbibliography,prb,10pt]{revtex4-2}
\usepackage[colorlinks=true,allcolors=blue,hypertexnames=false]{hyperref}
\usepackage{amsmath}
\usepackage{dsfont}
\usepackage{textcomp}
\usepackage{tabularx}
\usepackage{graphicx}
\usepackage{soul}
\usepackage{feynmp-auto}
\usepackage{footmisc}
\usepackage[export]{adjustbox}
\usepackage{wrapfig}
\usepackage{bbm}

\makeatletter
\def\@bibdataout@aps{%
 \immediate\write\@bibdataout{%
  @CONTROL{%
   apsrev42Control%
   \longbibliography@sw{%
    ,author="48",editor="1",pages="0",title="0",year="1"%
   }{%
    ,author="48",editor="1",pages="0",title="",year="1"%
   }%
  }%
 }%
 \if@filesw
  \immediate\write\@auxout{\string\citation{apsrev42Control}}%
 \fi
}%
\makeatother

\newcommand{\beg}{\begin{equation}}
\newcommand{\en}{\end{equation}}
\newcommand{\bp}{\mathbf p}
\newcommand{\bq}{\mathbf q}

\newcommand{\bk}{\mathbf k}
\newcommand{\br}{\mathbf r}
\newcommand{\bR}{\mathbf R}

\newcommand \bel  {\begin{align}}
\newcommand \enl  {\end{align}}

\newcommand{\eps}{\epsilon}
\newcommand{\veps}{\varepsilon}

\newcommand{\up}{\uparrow}
\newcommand{\dn}{\downarrow}
\newcommand{\dg}{^\dagger}

\newcommand{\Tr}{\mathrm{Tr}\,}

\newcommand{\ttau}[1]{\hat\tau_{#1}}

\newcommand{\hs}{\hat\sigma}

\def\XXint#1#2#3{{\setbox0=\hbox{$#1{#2#3}{\int}$}
     \vcenter{\hbox{$#2#3$}}\kern-.5\wd0}}

\begin{document}

\title{Odd-parity surface magnetism in a Kondo insulator}

\author{Maxim Dzero}
\affiliation{Department of Physics, Kent State University, Kent, Ohio 44242, USA}

\author{Brian M. Andersen}
\affiliation{Niels Bohr Institute, University of Copenhagen, DK-2100 Copenhagen, Denmark}

\date{\today}

\begin{abstract} 
We formulate a theory of spin-polarized tunneling between a nanowire tip of a topological Kondo insulator SmB$_6$ and the bi-collinear antiferromagnet Fe$_{1+x}$Te in which the tunneling conductance which measures magnetic contrast changes sign when the bias is reversed. First we show that no time-reversal-symmetric tip, however strong its spin–momentum locking is, can produce experimentally observed magnetic contrast, which implies that time reversal must be broken at the tip surface. Within a mean-field treatment of the surface theory of SmB$_6$ we find that the interaction between conduction $d$- and $f$-electrons drives a surface phase transition near 
$T_c\approx10$ K in which the time-reversal-odd component of the Kondo hybridization amplitude condenses in the same symmetry channel as an out-of-plane surface magnetization, gapping out the Dirac surface states. Below the transition the spin-resolved spectral function of each gapped surface Dirac cone is odd
in energy about its node. For a charge-neutral apex the Fermi level lies between the surface $\overline{\Gamma}$
and $\overline{X}$ Dirac nodes, so the spin-polarized conductance, which is determined by this function at the
bias energy, reverses its sign close to zero bias. This sign change is thus a spectroscopic
signature of time-reversal symmetry breaking on the surface of SmB$_6$. 
\end{abstract}

\maketitle
\section{Introduction}
In materials containing elements with partially filled $f$-shells, the
hybridization between the localized $f$-electrons and the conduction
electrons of the more extended $p$- and $d$-orbitals produces a coherent
many-body state below a characteristic temperature~\cite{Coleman2015}.
Depending on the total electron count per unit cell, this state is either a
heavy-fermion metal, with a strongly enhanced quasiparticle mass, or a Kondo
insulator, with a hybridization gap in the spectrum, SmB$_6$ being the
canonical example of the latter~\cite{Menth1969,Riseborough2000,Dzero2016}. In either case the
formation of the coherent state is a crossover rather than a phase
transition: no symmetry is broken as the low-temperature state develops, and
the sharp transition found in the slave-boson mean-field
theory~\cite{Read1983,Coleman1987} is an artifact of the approximation that disappears
once fluctuations are included~\cite{Millis1987}. It is also worth noting that even in the fractionalized scenario, where
the onset of Kondo screening becomes a genuine quantum phase
transition~\cite{Senthil2003,Senthil2004}, no symmetry is broken.

Recent spin-polarized tunneling measurements with SmB$_6$ nanowire tips on
the bicollinear antiferromagnet Fe$_{1+x}$Te~\cite{Aishwarya2022,Banerjee2025}
observed that the magnetic contrast, which appears as the stripe pattern of
the antiferromagnet in topographic images, changes sign when the bias is
reversed, and Banerjee \emph{et al.}~\cite{Banerjee2025} proposed that
time-reversal symmetry must be broken at the surface of SmB$_6$ below a phase
transition near $10$~K. This is all the more surprising because SmB$_6$ is a mixed-valent
compound: the average samarium valence is close to $2.6$, and as a consequence the samarium moments are
quenched by the valence fluctuations, so that at ambient pressure the
material shows no tendency toward magnetic order of any
kind~\cite{Cohen1970,Chazalviel1976,Cooley1995,Mizumaki2009}. Only hydrostatic pressure
on the scale of $10$~GPa, which closes the insulating gap and drives the
valence toward Sm$^{3+}$ while still leaving it intermediate, induces long-range
magnetic order~\cite{Barla2005,Derr2008,Butch2016}.

It can be shown quite generally (i.e.\ without making reference to any
specific microscopic model) that the bias reversal of the tunneling current
requires time-reversal symmetry breaking (TRSB) on the surface of the
SmB$_6$ tip. The question is then what drives this phase transition. In
this paper we address precisely this question by formulating a microscopic
model based on the orbital structure of the samarium $5d$ and $4f$
states~\cite{Takimoto2011,Alexandrov2013,Roy2014}. The quantity which
plays the central role in our theory is a spin scalar $\Delta_{PT}$, the time-reversal-odd
component of the same odd-parity $d$--$f$ hybridization that forms the bulk
gap. It is odd under time reversal simply because it is the imaginary part
of the hybridization amplitude. Such a term was first introduced for
SmB$_6$ in Ref.~\cite{Roy2014} as a Dirac mass which is odd under both
parity and time reversal, and more recently it was associated with the
imaginary part of the surface hybridization~\cite{Banerjee2025}. We note
that under the $C_{4v}$ point group of the (001) surface $\Delta_{PT}$
transforms according to the same one-dimensional representation as the
magnetization $M_z$ along the surface normal. Consequently, $\Delta_{PT}$
and $M_z$ cannot appear independently of each other: they are two
projections of one and the same Ising order parameter, so that a surface
with nonzero $\Delta_{PT}$ necessarily has a nonzero magnetic moment $M_z$. 
This is what we define here as odd-parity magnetism -- a magnetization carried by the
parity-odd order parameter of the $d$--$f$ mixing rather than by ordered
local moments. Order parameters of this kind, odd under both inversion and
time reversal, are the magnetic multipoles that give rise to a linear
magnetoelectric response~\cite{Watanabe2018,Hayami2018}. The order parameter
we propose should not be confused with the one in $p$-wave magnets of
Ref.~\cite{Hellenes2023}, which are compensated magnets with a spin
splitting odd in momentum and which preserve the combination of time reversal and
a lattice translation.

The question remains as to what interaction is responsible for the
condensation of $\Delta_{PT}$. It is well known that the extended Anderson
lattice model description of a Kondo insulator contains both the on-site
Coulomb repulsion $U_{\rm ff}$ of the $f$ shell and the interaction
$U_{\rm df}$ between the $f$ electrons and the conduction electrons on the
same site~\cite{Falicov1969,Riseborough2000,Onishi2000,Phan2010}. We propose
that $U_{\rm df}$ is the interaction that drives the surface TRSB
transition since after the Fock mean-field decoupling it becomes attractive in the $d$--$f$
excitonic channel. When we take into account the exchange interaction
$I_{\rm ff}$ within the $f$ shell as well, the mean-field theory predicts a
phase transition in which $\Delta_{PT}$ and the exchange (Weiss) field $h_z$
acting on the $f$ electrons appear together. A nonzero $h_z$ in turn produces
the surface magnetization $M_z\sim h_z/I_{\rm ff}$. We also find that most
of the contribution to the instability comes from the bulk states. In fact,
the bulk crystal is only a few percent away from its own instability, and
we include the band-bending potential in order to keep the ordered state
confined to the surface of the tip. For $U_{\rm df}$ of the order of
$100$~meV we obtain a transition temperature close to $10$~K, in agreement
with the temperature above which the magnetic contrast disappears in the
experiment. 

Below $T_c$ the order parameter acts on each of the three surface Dirac
cones as a mass term, and the masses of the cone at $\bar\Gamma$ and of the
cones at $\bar X$ and $\bar Y$ turn out to have opposite signs. The
spin-resolved spectral function of a single massive cone is an odd function
of energy with respect to the position of its node, and the competition
between cones with masses of opposite signs makes the spin-polarized
conductance change its sign within about $1$~meV of zero bias. We thus
obtain the bias reversal observed experimentally, and the theory determines
its position, magnitude and shape, which are all functions of the
microscopic model parameters.

Our paper is organized as follows. In Section~II we present the
model-independent argument that the magnetic contrast requires a
time-reversal-odd equilibrium state of the tip and identify the symmetry
channel of the order parameter. Section~III introduces the microscopic model
of the semi-infinite tip and its mean-field theory. In Section~IV we map the
instability in the plane of the two couplings, compute the layer profiles
and the temperature dependence of the order parameter, and obtain the
tunneling conductance. Section~V discusses the implications of our results,
including their bearing on the nature of the surface states of SmB$_6$, and
Section~VI contains our conclusions. Technical details are relegated to the
appendices.

\section{Qualitative discussion}
It is first natural to ask whether the bias reversal could come from the spin structure of the
tip's surface states alone, i.e. from the spin--momentum locking of a topological surface state, a
Rashba-split trivial band, or any other texture that assigns a definite spin to each
momentum, without the tip carrying a net moment. Initially, this was the interpretation proposed when the effect was discovered~\cite{Aishwarya2022}, and it
has since been applied to other topological tips~\cite{Wang2025}. Banerjee \emph{et al.} ~\cite{Banerjee2025} noted that
for a point-like apex the momentum sum removes the spin texture from the local density of
states. Below we will generalize the argument of Ref.~\cite{Banerjee2025} and show that bias reversal cannot come from the spin-structure of the surface states on a tip, for a reason that does not
depend on whether the bands on the tip are topological or not. Our argument below holds for any apex geometry and for interacting
electrodes.

Let us take the apex of the tip to be situated above a substrate at point $\mathbf r$. To lowest
order in the tunnelling matrix $\hat{T}_{\mathbf r}$, which maps tip states onto substrate states (see Appendix \ref{AppendixA}) the tunneling current is:
\begin{equation}
\begin{aligned}
I(V,\mathbf r)\;\propto\;&\int d\omega\,\big[f(\omega+eV)-f(\omega)\big]\,
\\&\times\mathrm{Tr}\big[\hat{\cal A}_{\mathbf r}(\omega+eV)\,\hat{A}_s(\omega)\big],
\label{eq:nogo-current}
\end{aligned}
\end{equation}
where $\hat{\cal A}_{\mathbf r}\equiv \hat{T}_{\mathbf r}\hat{A}_t\hat{T}_{\mathbf r}^{\dagger}$, $\hat{A}_{t,s}=\tfrac{i}{2\pi}\big(\hat{G}^R_{t,s}-\hat{G}^A_{t,s}\big)$ are the spectral matrices of tip and substrate, each measured from its own chemical
potential, so that the argument of $\hat{A}_t$ and the tip occupation carry the same shift $eV$ and $\hat{G}_{t,s}^{R(A)}$ are the corresponding retarded and advanced matrix propagators.
In Eq. \eqref{eq:nogo-current} the trace runs over the substrate's sites and spin. The tip enters only through the
Hermitian matrix $\hat{\cal A}_{\mathbf r}(\omega)$, which represents its spectral density as seen from the substrate through
the barrier.

For spin-$1/2$ (or an $f$-electron Kramers doublet), time reversal on either electrode is the anti-unitary operator
$\Theta=UK$, with $K$ complex conjugation in the site--spin basis and $U$ acting as
$i\sigma_y$ on every site. So $\Theta$ acts on a matrix block by
block: the $2\times2$ spin block between sites $\mathbf r$ and $\mathbf r'$ transforms as
$M(\mathbf r,\mathbf r')\mapsto\sigma_yM^{*}(\mathbf r,\mathbf r')\sigma_y$. Every Hermitian
matrix splits uniquely into a time-reversal-even and a time-reversal-odd part $M=M^{\rm even}+M^{\rm odd}$ where  $M^{\rm even/odd}$ are defined according to
\begin{equation}
M^{\rm even/odd}=\tfrac12\big(M\pm\Theta M\Theta^{-1}\big),
\label{eq:nogo-split}
\end{equation}
so that  $\Theta M^{\rm even}\Theta^{-1}=M^{\rm even}$ and
$\Theta M^{\rm odd}\Theta^{-1}=-M^{\rm odd}$ and we have omitted the dependence of $M$ on $\br$ and $\br'$ for brevity. Writing each block as
$M=a_0\sigma_0+\mathbf a\cdot\boldsymbol\sigma$ and using
$\sigma_y\boldsymbol\sigma^{*}\sigma_y=-\boldsymbol\sigma$, we have
\begin{equation}\label{eq:nogo-components}
\begin{aligned}
&M^{\rm even}=\mathrm{Re}(a_0)\,\sigma_0
+i\,\mathrm{Im}(\mathbf a)\cdot\boldsymbol\sigma,\\
&M^{\rm odd}=i\,\mathrm{Im}(a_0)\sigma_0
+\mathrm{Re}(\mathbf a)\cdot\boldsymbol\sigma,
\end{aligned}
\end{equation}
On a single site $\br=\br'$ $M$ is hermitian which makes $a_0$ and $\mathbf a$ real, so the on-site block of
$M^{\rm odd}$ is a spin density, $\mathbf a\cdot\boldsymbol\sigma$, and
the on-site block of $M^{\rm even}$ is proportional to the unit matrix.  Between different
sites $M^{\rm odd}$ may also contain orbital currents, $i\,\mathrm{Im}\,a_0$. For a spectral
matrix, $A^{\rm odd}$ is therefore its magnetic part: the spin-resolved density of states,
together with any orbital-current contribution. 

The spin texture of a nonmagnetic band is not part of $M^{\rm odd}$. For such a band with ${A}(\mathbf k)=a_0(\mathbf k)\sigma_0+\mathbf a(\mathbf k)
\cdot\boldsymbol\sigma$ time reversal invariance of ${A}(\mathbf k)$ requires that $a_0(\bk)=a_0(-\bk)$ and $\mathbf a(-\mathbf k)=-\mathbf a(\mathbf k)$.
The Fourier transform of a real odd function is imaginary and vanishes at zero separation, $\br=\br'$, so
in real space the texture is an imaginary spin-dependent amplitude between different sites,
$i\,\mathrm{Im}\,\mathbf a(\mathbf r,\mathbf r')\cdot\boldsymbol\sigma$ with
$\mathbf r\neq\mathbf r'$, and comparing it to \eqref{eq:nogo-components} we see that it belongs to $M^{\rm even}$. Spin--momentum locking, no matter how
strong, is therefore does not contribute to spin density.

In the ordered state of Fe$_{1+x}$Te, which we assume to be perfectly commensurate and collinear, 
reversing every moment is the same as translating the magnetic pattern by half its period
$\mathbf R_{\frac{1}{2}}$, so that $\mathbf Q\cdot\mathbf R_{\frac{1}{2}}=\pi$ and also
$\Theta A_s\Theta^{-1}={\cal T} A_s{\cal T}^{-1}$, with ${\cal T}$ the translation by $\mathbf R_{\frac{1}{2}}$.
Hence $A_s^{\rm even}$ is invariant under ${\cal T}$ and $A_s^{\rm odd}$ changes sign. Moving the
tip by $\mathbf R_{{1}/{2}}$, $\hat{\cal A}_{\mathbf r+\mathbf R_{{1}/{2}}}={\cal T} \hat{\cal A}_{\mathbf r}{\cal T}^{-1}$, leaves
the contribution of $A_s^{\rm even}$ to Eq.~(\ref{eq:nogo-current}) unchanged and reverses
the sign of $A_s^{\rm odd}$. The magnetic contrast, defined as the part of the signal that changes
sign when the tip is moved by half a magnetic period is therefore determined exactly by $A_s^{\rm odd}$. Introducing the magnetic contrast current $\delta I_{\br,\br+\bR_{{1}/{2}}}^{\rm odd}(V)=\big[I(V,\mathbf r)-I(V,\mathbf r+\mathbf R_{{1}/{2}})\big]/2$ we have 
\begin{equation}
\begin{aligned}
\delta I_{\br,\br+\bR_{{1}/{2}}}^{\rm odd}(V)
\propto&\int d\omega\,\big[f(\omega+eV)-f(\omega)\big]\\&\times
\mathrm{Tr}\big[{\cal A}_{\mathbf r}(\omega+eV)\,A_s^{\rm odd}(\omega)\big].
\end{aligned}
\label{eq:nogo-contrast}
\end{equation}
Consequently, the contribution from $A_s^{\rm even}(\omega)$, which accounts for the charge density of states and any time-reversal-even
spin--orbit texture of the substrate can also be obtained by computing $\delta I_{\br,\br+\bR_{{1}/{2}}}^{\rm even}(V)=\big[I(V,\mathbf r)+I(V,\mathbf r+\mathbf R_{{1}/{2}})\big]/2$. 

At a
point-like apex only the local block enters,
$A_s(\mathbf r,\mathbf r;\omega)=\rho_0(\omega)\sigma_0+\rho_1(\omega)
\cos(\mathbf Q\cdot\mathbf r+\varphi_0)\,(\hat{\mathbf n}\cdot\boldsymbol\sigma)$, where $\rho_0$ is the substrate's local density of states per spin,
$\rho_1$ the amplitude of its spin polarization, and $\hat{\mathbf n}$ the direction of the
Fe moments. The time-reversal odd part of $A_s(\mathbf r,\mathbf r;\omega)$ is the second ($\propto\rho_1(\omega)$) term. The integrand of Eq.~(\ref{eq:nogo-contrast}) is then
\beg\label{Eq4Integrand}
\begin{aligned}
\mathrm{Tr}\big[{\cal A}_{\mathbf r}(\omega+eV)\,A_s^{\rm odd}(\omega)\big]&=\rho_1(\omega)\cos(\mathbf Q\cdot\mathbf r+\varphi_0)\\&\times\,\rho_{\rm spin}(\omega+eV), 
\end{aligned}
\en
where
$\rho_{\rm spin}(\omega)\equiv\mathrm{Tr}[(\hat{\mathbf n}\cdot\boldsymbol\sigma){\cal A}_\br]$ is the spin-resolved
spectral function of the apex, which the spin-polarized conductance reads at the bias energy.
If the tip's equilibrium state is time-reversal symmetric,
then $\Theta G^R_t(\omega)\Theta^{-1}=G^A_t(\omega)$ and hence
$\Theta A_t\Theta^{-1}=A_t$: the tip's spectral matrix has no odd part, whatever its band
structure, spin--orbit coupling or interactions are. If, in addition, the barrier contains no
magnetic field or magnetic impurities, then $\Theta T_{\mathbf r}\Theta^{-1}=T_{\mathbf r}$, and
${\cal A}_{\mathbf r}$ remains time-reversal invariant for every apex geometry.

We can now use the expressions above to prove that $\delta I_{\br,\br+\bR_{{1}/{2}}}^{\rm odd}(V)$ must necessarily be zero when the tip is time reversal invariant. 
For antiunitary operator $\Theta=UK$ and any matrix $M$ we have
$\mathrm{Tr}[\Theta M\Theta^{-1}]=\mathrm{Tr}[UM^{*}U^{\dagger}]=\mathrm{Tr}\,M^{*}
=(\mathrm{Tr}\,M)^{*}$. Let ${\cal C}\equiv\mathrm{Tr}[{\cal A}_\br A_s^{\rm odd}]$ be the
integrand of Eq.~(\ref{eq:nogo-contrast}) at one energy. Time reversal gives
\begin{equation}\label{eq:jimag}
{\cal C}^{*}=\mathrm{Tr}\big[(\Theta {\cal A}_\br\Theta^{-1})(\Theta A_s^{\rm odd}\Theta^{-1})\big]=-{\cal C},
\end{equation}
so ${\cal C}$ is purely imaginary. At the same time since both ${\cal A}_\br$ and $A_s^{\rm odd}$ are hermitian, it follows
\beg\label{Creal}
{\cal C}^{*}=\mathrm{Tr}[({\cal A}_\br A_s^{\rm odd})^{\dagger}]={\cal C}
\en
and it appears that ${\cal C}$ must be real which clearly contradicts Eq. \eqref{eq:jimag}.
Therefore ${\cal C}$ is zero at every energy and every tip position, and the magnetic
contrast $\rho_{\rm spin}(\omega)$ vanishes at every bias and temperature. 

The nonzero magnetic contrast therefore requires a time-reversal-odd component in the equilibrium
state of the tip apex, and the question becomes what that component is and how it enters into the expression for 
$\rho_{\rm spin}(\omega)$. Banerjee \emph{et al.}~\cite{Banerjee2025} identify it with the
imaginary part of a complex surface $d$--$f$ hybridization ${\cal V}=V_{\rm h}+i\Delta$. In the
eight-band description of SmB$_6$ \cite{Takimoto2011,Alexandrov2013,Dzero2016} nonzero imaginary part of ${\cal V}$ corresponds to the expectation value of the operator
\beg\label{eq:DPT}
\hat{\cal M}=\left(\,\ttau{2}\otimes\hs_0\otimes\hat{\mathbbm 1}_2\right)\Delta_{PT},
\en
where $\hat{\tau}_2$ acts between four component $d$- and $f$ spinors \cite{Roy2014}, $\hat{\sigma}_0$ acts in spin subspace while $\hat{\mathbbm 1}_2$ acts in intra oribital subspace of $d$- and $f$-electron states \cite{Roy2014}:
\beg
\Psi\dg\big(\ttau{2}\otimes\hs_0\otimes\hat {\mathbbm 1}_2\big)\Psi=-i\sum_{\alpha\sigma}\big(d\dg_{\alpha\sigma} f_{\alpha\sigma}-f\dg_{\alpha\sigma} d_{\alpha\sigma}\big).
\en
Operator $\hat{\cal M}$ describes the parity- and time-reversal-odd Dirac mass which was first introduced for SmB$_6$ in
Ref.~\cite{Roy2014}. Since the Kondo-singlet state of this form could be favored by the interaction
between the $d$- and $f$-electrons we will adopt the order parameter \eqref{eq:DPT} in what follows. 

Let us consider for definiteness the (001) surface of the SmB$_6$ tip, whose point group is
$C_{4v}$. We introduce $\Delta_{PT}$ as the real coefficient of the Hermitian operator in
Eq.~\eqref{eq:DPT}. Under the symmetry operations of the surface it transforms according
to the one-dimensional representation $A_2$: it is invariant under the rotations about the
surface normal and changes sign under reflections in each of the four vertical mirror
planes, the sign under the mirrors coming from the opposite parities of the $d$- and
$f$-states. Being a single real component odd under time reversal, $\Delta_{PT}$ is an
Ising order parameter, with two domains $\pm\Delta_{PT}$ related by time reversal.
Without a training field each cool-down selects one of the two domains at
random, as anticipated in Ref.~\cite{Banerjee2025}, where it was also proposed
that the domain can be trained by $\mathbf E\cdot\mathbf B$.

The operator in Eq.~\eqref{eq:DPT} is, however, a spin scalar, whereas the
magnetic contrast, Eq.~\eqref{Eq4Integrand}, reads the spin density of the
apex. A spin-scalar order parameter can therefore enter the contrast only
through the spin polarization it induces in the surface states, i.e.\
through a finite magnetization of the surface, which we need to identify. The magnetization of the tip surface along its normal is the expectation value of the following operator
\beg
\hat M_z=\frac{\mu_B}{2}\sum_{\mathbf R\in{\rm surf}}
\Psi\dg_{\mathbf R}\big[\big(g_d\hat {\cal P}_d+g_f\hat {\cal P}_f\big)\otimes\hs_z\otimes\hat{\mathbbm 1}_2\big]\Psi_{\mathbf R},
\label{eq:Mz}
\en
where $\hat{\cal P}_{d,f}=\big(\ttau{0}\pm\ttau{3}\big)/2$, $g_d$ and $g_f$ are the effective $g$-factors of the $d$- and $f$-orbital quartets and the sum
runs over the sites of the outermost layer. Since
$g_d\hat {\cal P}_d+g_f\hat {\cal P}_f=\tfrac12(g_d+g_f)\ttau{0}+\tfrac12(g_d-g_f)\ttau{3}$, $\hat M_z$ is
a combination of the two operators $\ttau{0}\otimes\hs_z\otimes\hat{\mathbbm 1}_2$ and
$\ttau{3}\otimes\hs_z\otimes\hat{\mathbbm 1}_2$. 

Both $\Delta_{PT}$ and $M_z$ are connected by symmetry. The magnetization of the
surface layers along the surface normal, $M_z$, is odd under time reversal
and also transforms according to $A_2$: a moment along the normal is
unchanged by rotations about it and, being an axial vector, is reversed by a
reflection in any plane that contains it. The two quantities therefore lie
in one symmetry channel, and a term $\propto\Delta_{PT}M_z$ is allowed in the
free energy, so that as soon as one of them becomes nonzero the other is
induced. They are two components of one order parameter rather than
competing orders, and the spin density read by the substrate is carried by
the magnetization $M_z$ that $\Delta_{PT}$ induces. A static moment of this
kind, of order $10^{-2}\mu_B$, is consistent with the zero-bias intercept of
the data analysed in Ref.~\cite{Banerjee2025}.

We note that the time-reversal-even hybridization, of magnitude $V_{\rm h}$, is generated
by the Hamiltonian at every temperature and only crosses over, whereas $\Delta_{PT}$ is
forbidden above $T_c$ by time reversal and can only condense below it. The transition is
therefore a transition in the $A_2$ channel and not in $V_{\rm h}$, which enters as a fixed
background that sets the stiffness of the ordering field, $\propto V_{\rm h}^2+\Delta_{PT}^2$
in the notation of Ref.~\cite{Banerjee2025}. This is also why $T_c$ cannot exceed the scale
on which $V_{\rm h}$ itself develops.

What such an order parameter does to the apex spectrum follows from the surface states
alone. Each Dirac cone of the (001) surface
\beg\label{hj}
h_j(\bq)=\eps_j +v_{Fj}\big(q_y\hs_x-q_x\hs_y\big)+m_j\hs_z,
\en
where $\eps_j$ ($j=1,2,3$) is a node energy, admits a single
on-site mass that is odd under time reversal. A time-reversal-odd order parameter in the $A_2$ irreducible representation therefore acts on every cone as a
Dirac mass. Since $\Delta_{PT}$ and $M_z$ belong to this channel, both will contribute
\beg
m_a=w_\Delta(a)\,\Delta_{PT}+w_M(a)\,h_z,
\label{eq:mj}
\en
where $h_z$ is the exchange field produced on the surface electrons by the induced
magnetization, and the weights $w_\Delta(a)$ and $w_M(a)$ are the matrix elements of the
operators in Eqs.~\eqref{eq:DPT} and \eqref{eq:Mz} between the surface states of cone $a=(\overline{\Gamma},\overline{X},\overline{Y})$.
Their values, and the relative sign of $\Delta_{PT}$ and $h_z$, which decides whether the two
contributions add or partially cancel on a given cone, follow from the microscopic
calculation of Section \ref{SectionIV}.

The 
eigenstates for each Dirac cone are $\eps_j\pm E_j$, where $E_j=(v_j^2q^2+m_j^2)^{1/2}$. The eigenvectors carry the spin projection
$\pm m_j/E_j$ along the mass axis, while the density of states of the cone is $E_j/2\pi v_j^2$. These
two factors combine into a spin-resolved local density of states (up to the tunneling weight of the cone at the apex) that is odd relative to the position of the node (see Appendix \ref{AppendixB} for details):
\beg
\rho_{\rm spin}(\omega)=\sum\limits_{j=1}^3\frac{m_j}{2\pi v_j^2}\;{\rm sgn}(\omega-\eps_j)\;
\theta\big(|\omega-\eps_j|-|m_j|\big),
\label{eq:oddstep}
\en
so the upper band is polarized one way,
the lower band the other, and no spin weight sits inside the gap. Inserted into Eq.~(\ref{eq:nogo-contrast}) with a substrate polarization
$\rho_1$ that is smooth over the bias window, Eq.~(\ref{eq:oddstep}) gives
for a single cone a spin-polarized conductance proportional to
$\rho_{\rm spin}(eV)$, which reverses its sign at $eV=\eps_j$, and a
magnetic contrast current $\delta I_{\br,\br+\bR_{{1}/{2}}}^{\rm odd}(V)$
that is symmetric about $eV=\eps_j$. For a single cone the reversal bias is
thus set by the offset of the node from the Fermi level, not by the width of
any resonance. On the (001) surface of SmB$_6$ there are three cones, and the
contributions of the other cones keep $\rho_{\rm spin}(\omega)$ finite even
where $|\omega-\eps_j|<|m_j|$ for a given cone. As we show in
Section~IV.C, the masses of the cones have opposite signs, and the position
of the reversal is then decided by the competition between them.

\section{Model}
There are several important issues which need to be resolved so that the ideas presented above can be put on a firm footing. First, the weights $m_j/v_j^2$  in Eq. \eqref{eq:oddstep}, the node offsets $\eps_j$, and the current-voltage line shape once the Dirac cone is dressed by the
hybridization all require the quantitative description within the concrete microscopic model. The same holds for Eq.~\eqref{eq:oddstep} which fixes the sign structure of
$\rho_{\rm spin}(\omega)$ and hence the existence of the reversal, but its position, magnitude and
shape must also be computed from a microscopic theory. Also the ratio $\delta I_{\br,\br+\bR_{{1}/{2}}}^{\rm odd}/\delta I_{\br,\br+\bR_{{1}/{2}}}^{\rm even}$
is an experimental observable and it is one of the quantities which microscopic calculation can provide. We will address these issues in what follows using the interacting surface theory of SmB$_6$.

\subsection{Model Hamiltonian in the bulk}
In the first part of this Section we will follow the avenue of Ref. \cite{Roy2014}. Specifically, in order to describe the bulk states in SmB$_6$ we consider the following mean-field Hamiltonian 
\begin{equation}\label{Hbulk}
H_{\rm bulk}(\bk)=\left(  \begin{array}{c c}
H^d(\bk) & V_h(\bk) \\
V^\dagger_h(\bk) & H^f(\bk)
\end{array} \right),
\end{equation} 
where $H^d(\bk)$, $H^f(\bk)$ and $V_h(\bk)$ are 4-dimensional matrices which act in the basis $\Psi=[\Psi_d,\Psi_f]^T$ defined by
\begin{equation}\label{basis}
\Psi_l=\left[l_{1\uparrow}, l_{1\downarrow}, l_{2\uparrow}, l_{2\downarrow} \right]^T
\end{equation}  
with $l=d,f$ and $f_{j\up,\dn}$ refer to the components of the $j$-th Kramers doublet. The diagonal blocks in \eqref{Hbulk} $H^l(\bk)$ are given by 
\begin{equation}\label{tightbinding}
\begin{aligned}
H^l(\bk)&=\epsilon_l \hat{{\mathbbm 1}}_4\\&+ t_l \left( \begin{array}{c c}
\hat{\phi}_1(\bk)+\eta_l \hat{\phi}_2(\bk) & (1-\eta_l) \hat{\phi}_3 (\bk) \\
(1-\eta_l) \hat{\phi}_3 (\bk) &  \eta_l \hat{\phi}_1(\bk)+ \hat{\phi}_2(\bk)
\end{array} 
\right),
\end{aligned}
\end{equation}
where $t_{d}$ and $t_{f}$ the are hopping amplitudes, and $\epsilon_{d}$ and $\epsilon_{f}$ are the corresponding chemical potentials, for the $d$- and $f$-electrons, correspondingly. In the expression \eqref{tightbinding} $\hat{\phi}_j(\bk)=\hat{\sigma}_0 \phi_j(\bk)$, where $\hat{\sigma}_0$ and $\hat{{\mathbbm 1}}_4$ are respectively the two- and four-dimensional identity matrices and functions $\phi_j(\bk)$ are 
\begin{equation}\label{phis}
\begin{split}
&\phi_1(\bk)=\frac{1}{2} \left( c_x + c_y + 4 c_z \right), \\
&\phi_2(\bk)=\frac{3}{2} \left( c_x+c_y\right), \: \: \phi_3(\bk)=\frac{\sqrt{3}}{2} (c_x-c_y) \\
\end{split}
\end{equation}
with $c_\alpha=\cos{k_\alpha}$, for $\alpha=x,y,z$ (in what follows next we choose the units in which the lattice spacing $a=1$). The hybridization matrix reads as 
\begin{equation}
V_h(\bk)= \frac{V}{4}\left( \begin{array}{c c}
3 (\bar{\sigma}_x+\bar{\sigma}_y) & -\sqrt{3} (\bar{\sigma}_x-\bar{\sigma}_y) \\
-\sqrt{3} (\bar{\sigma}_x-\bar{\sigma}_y) & \bar{\sigma}_x+\bar{\sigma}_y+4 \bar{\sigma}_z
\end{array} 
\right),
\end{equation}
where $\bar{\sigma}_\alpha=\hat{\sigma}_\alpha \sin{k_\alpha}$ for $\alpha=x,y,z$, and $\hat{\sigma}_{x,y,z}$ are the standard two-dimensional Pauli matrices. The bare hybridization amplitude is represented by $V$. In this work we restrict ourselves with hole-like $f$-states, i.e., $t^d t^f<0$, only for which a topologically non-trivial insulating state emerges below the Kondo coherence temperatures, which for SmB$_6$ is $\sim 50$ K \cite{Dzero2010,DMFT2013,Sigrist2014}. The values of the model parameters $V$, $t_{d,f}$, $\eta_{d,f}$ and $\eps_{d,f}$ are chosen in a way so that the resulting band structure reproduces the one measured experimentally \cite{Alexandrov2013,Roy2014,Dzero2016}.

\subsection{Layer Hamiltonian}
Let us now turn our discussion to the tunneling problem. Mainly for simplicity we are going to assume that the apex of the tip is a (001) surface. The surface states which develop time-reversal symmetry breaking instability discussed in Section~II will have to be 
obtained by terminating the bulk crystal rather than by projecting it. This is because the surface Dirac cones are not well separated from the bulk bands. Ignoring this would severely underestimate the strength of the effect because, as we will show below, most of surface instability is actually governed by the bulk states. 

In order to set the stage we start with the noninteracting case first. Given that 
$k_z$-dependence of $H_{\rm bulk}(\bk)$ enters through $\cos k_z$ and $\sin k_z$ only, we write it as
\beg
\hat{H}_{\rm bulk}(\bk_{\parallel},k_z)=\hat h(\bk_\parallel)+\big[\hat\rho\,e^{ik_z}+\hat\rho\dg\,e^{-ik_z}\big].
\label{eq:m-hrho}
\en
Here $\hat h$ and $\hat\rho$ the intra- and inter-layer $8\times8$ blocks which we define as follows:
\beg\label{hkparrho}
\begin{aligned}
\hat{h}(\bk_\parallel)&=\frac{1}{2}\left[\hat{H}_{\rm bulk}(\bk_\parallel,0)+\hat{H}_{\rm bulk}(\bk_{\parallel},\pi)\right], \\
\hat{\rho}&=\frac{1}{4}\left[\hat{H}_{\rm bulk}(\bk_\parallel,0)-\hat{H}_{\rm bulk}(\bk_{\parallel},\pi)\right]\\&+\frac{i}{2}\left[\hat{h}(\bk_\parallel)-\hat{H}_{\rm bulk}(\bk_{\parallel},\pi/2)\right].
\end{aligned}
\en

As a next step we consider a slab of $N$ layers labelled by $j=0,\dots,N-1$ with in-plane momentum $\bk_\parallel=(k_x,k_y)$ conserved. The corresponding $8N\times 8N$ Hamiltonian for the slab can be written down using \eqref{eq:m-hrho}:
\beg
\begin{aligned}
H_{\rm slab}(\bk_\parallel)=\;&\sum_{j=0}^{N-1}\Big[\hat h(\bk_\parallel)+U(j)\Big]\otimes|j\rangle\langle j|\\
&+\sum_{j=0}^{N-2}\Big[\hat\rho\otimes|j\rangle\langle j{+}1|+{\rm h.c.}\Big].
\end{aligned}
\label{eq:m-slab}
\en
The only surface-specific input which we include here is the band-bending
potential $U(j)=U_{\rm b}e^{-j a/\lambda_B}$, diagonal in all internal indices and it was introduced in \cite{Roy2014}
to account for the surface charge which shifts the Dirac points relative to the Fermi
level. 

\begin{figure}
\includegraphics[width=0.95\linewidth]{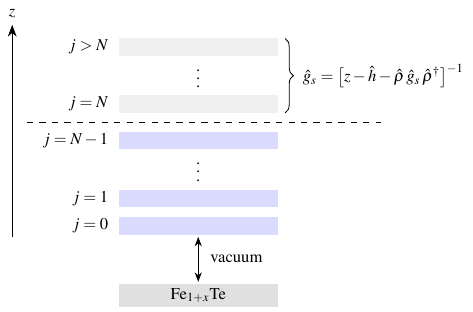}
\caption{Schematic representation of the tunneling contact. The Fe$_{1+x}$Te is located at $z<0$ and the SmB$_6$ tip occupies the region $z> 0$. The apex of the tip, $j=0$, faces the Fe$_{1+x}$Te substrate across a
vacuum gap at $z=0$. Time reversal breaking happens for $0\leq j\leq N-1$, while $j\geq N$ represents the time-reversal
symmetric nonmagnetic bulk, which enters through its surface Green's function $\hat g_s$.}
\label{Fig-Schematic}
\end{figure}

The model Hamiltonian \eqref{eq:m-slab} describes a slab with two surfaces, at $j=0$ and $j=N-1$, which are equivalent
by inversion when $U_{\rm b}=0$. For the tunneling problem we need to retain only one surface  which is in contact with the substrate. 
The time-reversal breaking instability and the corresponding order parameter are localized at the surface and leak into the bulk, but the order parameter must vanish deep inside the bulk. All this physics is contained in the spectral function of the apex, $j=0$. What the calculation therefore requires is the single-particle propagator of the tip: the tunneling current, Eq.~\eqref{eq:IV}, and the magnetic contrast, Eq.~\eqref{Eq4Integrand}, contain the tip only through its spectral matrix $\hat A_t={i}\big(\hat G^R_t-\hat G^A_t\big)/{2\pi}$ at the apex. The self-consistency conditions in the mean-field approximation for the interacting theory also require the corresponding order parameter averages on the layers where the mean fields
are finite, and the linearized problem requires the susceptibility built from the blocks $\hat G_{jj'}^{R(A)}$ between the
same layers which are blocks of the propagator on the layers $j\leq N-1$. 

We therefore divide the
semi-infinite crystal into the explicit region $S=\{0,\dots,N-1\}$, whose Hamiltonian $\hat H_{\rm slab}(\bk_\parallel)$ is
Eq.~\eqref{eq:m-slab} and the nonmagnetic remainder $j\geq N$.
This requires $U(j)$ to be negligible for $j\geq N$, which fixes $N$. The two parts are coupled
only through $\hat\rho$ between the layers $N-1$ and $N$, so the propagator on $S$ is exactly
\beg\label{GS}
\hat G_{S}(\bk_\parallel,\omega)=\Big[\hat{\omega}-\hat H_{\rm slab}-\hat\Sigma_s\otimes|N{-}1\rangle\langle N{-}1|\Big]^{-1}
\en
at each $(\bk_\parallel,\omega)$, where $\hat\Sigma_s=\hat\rho\,\hat g_s\,\hat\rho\dg$ is the self-energy on the layer $N-1$,
$\hat{\omega}=(\omega+\veps_F\pm i0)\hat{\mathbbm 1}_{8N\times 8N}$ and $\veps_F$ is the Fermi energy. Here $\hat g_s$ is the propagator on the first layer of the nonmagnetic bulk
remainder ($j\geq N$) shown in Fig.~\ref{Fig-Schematic}. Removing the layer $N$ from the remainder leaves an
identical semi-infinite stack for which the matrix function $\hat g_s(\bk_\parallel,\omega)$ must be computed self-consistently 
\beg\label{gs}
\hat g_s(\bk_\parallel,\omega)=\Big[\hat{\omega}-\hat h(\bk_\parallel)-\hat\rho\,\hat g_s(\bk_\parallel,\omega)\,\hat\rho\dg\Big]^{-1}.
\en
Of the several solutions of this matrix equation, the physical one is the one whose imaginary part is non-positive for
${\rm Im}\,\omega>0$. With energies measured from the chemical potential of the tip, as in Sec.~II, $\omega+\veps_F+i0^+$ gives
the retarded propagator entering the tunneling spectra. Matrix function $\hat{g}_s(\bk_\parallel,\omega)$ can be computed using the decimation method of Refs. \cite{LopezSancho1984,LopezSancho1985}.

\begin{figure}
\includegraphics[width=\linewidth]{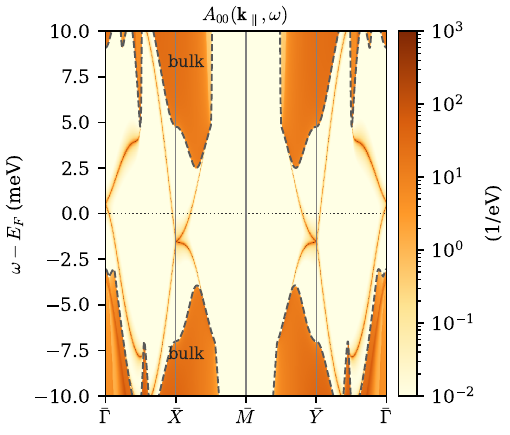}
\caption{Surface spectral function of the nonmagnetic apex of (001) SmB$_6$ tip,
$A_{00}(\bk_\parallel,\omega)$,
along $\bar\Gamma$--$\bar X$--$\bar M$--$\overline{Y}$--$\bar\Gamma$ of the surface Brillouin zone shown using the logarithmic color scale,
in units of eV$^{-1}$. Energies are measured from the Fermi level $E_F=-6.35$~meV (dotted line), fixed by
charge neutrality of the bulk. Dashed lines are the edges of the projected valence and conduction bulk bands,
$\max_{k_z}E_{\rm v}(\bk_\parallel,k_z)$ and $\min_{k_z}E_{\rm c}(\bk_\parallel,k_z)$ correspondingly. Parameters of
$\hat H_{\rm bulk}$: $t_d=-2$~eV, $t_f=0.05$~eV, $\eta_d=-0.3$, $\eta_f=-0.29$, $V=36.5$~meV, with
$\epsilon_d$ and $\epsilon_f$ chosen so that the $d$ and $f$ levels at $\Gamma$ lie at $0.2$~eV and
$-10$~meV ($\epsilon_d=4.4$~eV, $\epsilon_f=-116.5$~meV). The band-bending potential is
$U(j)=U_0e^{-j/\lambda_B}$ with $U_0=-1.5$~meV and $\lambda_B=2$ (units of lattice spacing). We treat $N=48$ layers explicitly
and the information about the remainder of the tip enters through $\hat g_s$, Eq.~\eqref{gs}. The surface
Dirac points lie at $+0.55$~meV ($\bar\Gamma$) and $-1.50$~meV ($\bar X$), inside the indirect bulk gap,
and the cones merge with the bulk continuum within a few meV of $E_F$.}
\label{Fig-A00}
\end{figure}

After $\hat{g}_s$ has been computed, we use it to evaluate \eqref{GS}. In Fig. \ref{Fig-A00} we show the results for the single-particle spectral function on the surface later $j=0$
\beg\label{A_00}
A_{00}(\bk_{\parallel},\omega)=-\frac{1}{\pi}\textrm{Im}\textrm{Tr}\left[\,\hat{G}_S(\bk_\parallel,\omega)\right]_{00}
\en
shown in a frequency region from approximately $-16\,{\rm meV}$ to $4\,{\rm meV}$. We note that from this plot it follows that the Dirac cones are within few meV from the Fermi energy , i.e. they are not well separated from the bulk bands. 

\subsection{Interaction and mean-field decoupling}
The Hamiltonian of Ref.~\cite{Roy2014} contains no interaction since it is obtained in the limit
of infinite repulsion $U_{\rm ff}$ between the $f$-electrons with the hybridization $V$ and the $f$-level
$\eps_f$ already renormalized by projecting out the doubly occupied states in the $f$-shell. Since we need to avoid the double counting we cannot 
decouple it a second time to generate $\Delta_{PT}$. We believe that the interaction
that can condense $\Delta_{PT}$ must be a different one, and given the mixed-valence nature of SmB$_6$ the natural candidate is the
interaction between the conduction and the $f$-electrons on the same site, which the
standard Anderson lattice construction does not include \cite{Dzero2010,DzeroLargeN,Roy2014,Dzero2016}:
\beg
H_{\rm df}=U_{\rm df}\sum_{{\bf R}\in{\rm S}}\,\sum_{\alpha\sigma}d\dg_{\alpha\sigma}d_{\alpha\sigma}\,
\sum_{\beta\sigma'}f\dg_{\beta\sigma'}f_{\beta\sigma'},
\label{eq:m-Udf}
\en
where we have restricted the summation to the layers which belong to region with $0\leq j\leq N-1$, Fig. \ref{Fig-Schematic}.
Decoupled in the exchange (Fock) channel, a repulsive $U_{\rm df}$ is attractive for
$d$--$f$ bilinears and the corresponding mean-field decoupled Fock term is
\beg\label{FockTerm}
-U_{\rm df}\sum\limits_{\alpha\sigma}\sum\limits_{\alpha'\sigma'}\langle d\dg_{\alpha\sigma}f_{\alpha'\sigma'}\rangle
f\dg_{\alpha'\sigma'}d_{\alpha\sigma}, 
\en
which is the mean-field form of an excitonic order
between the even-parity $d$- and odd-parity $f$-states~\cite{Halperin1968}. 

Two structural properties of the decoupling scheme \eqref{FockTerm} decide what kind of
order it can describe. First, a density--density interaction, as well as any interaction
which conserves $n_d$ and $n_f$ separately, is invariant under the relative phase rotation
$d\to e^{i\varphi}d$, $f\to f$. Since the energy gain is determined by $|\langle d\dg f\rangle|^2$,
it weighs the time-reversal-even and the time-reversal-odd components of the on-site mixing,
$\ttau{1}$ and $\ttau{2}$, with the same coupling, and likewise the spin-scalar and the
spin-dependent bi-linears. The interaction therefore supplies the magnitude of
$\langle d\dg f\rangle$ but cannot choose its phase or its spin structure. Second, the
single-particle Hamiltonian breaks the relative $U(1)$ explicitly through the hybridization,
which already fixes the phase of the $\ttau{1}$ spin-scalar component at the level of the
bulk mean-field theory: in that channel the interaction merely renormalizes $V$. Condensation
in any other channel breaks a symmetry that the Hamiltonian respects, time reversal in the
case of $\ttau{2}$, and is therefore a genuine second-order phase transition. In this regard this situation is similar to the composite pairing
theories in heavy-fermion materials \cite{Flint2008,Flint2010,Flint2011,Chandra2013}.
Since all
channels share the same coupling, the one that condenses first is the one with the largest
susceptibility ($\chi_{\tau_1}$ vs. $\chi_{\tau_2}$) and the comparison between channels does not depend on the value of $U_{\rm df}$.

% MD: this paragraph carries the "transverse channel" discussion of 24 Sep.
The symmetric phase which we referred to above is not a state without hybridization. The time-reversal-even
hybridization amplitude $V$ is present at temperatures below the Anderson lattice coherence temperature $T^*\simeq 50$ K, generated by the Hamiltonian itself, and the
Dirac cones of the surface exist because of it. Against this background the interaction can
act in two ways: increase the magnitude of the existing amplitude, the longitudinal channel
$\ttau{1}$, or rotate the amplitude into the time-reversal-odd direction, the transverse
channel $\ttau{2}$. On the surface cones the two are not equivalent: $\ttau{2}$ anticommutes
with the cone Hamiltonian and opens a gap at every node, so that the single-particle energy
gained by a condensate $\Delta_{PT}$ is that of a Dirac mass while $\ttau{1}$ commutes with it and
only displaces the cones. As a result, the energy gain for the condensation in  $\ttau{2}$ exceeds the one in the $\ttau{1}$ channel.
The transverse susceptibility $\chi_{\tau_2}$ therefore exceeds the
longitudinal one $\chi_{\tau_1}$. In passing we note that this ordering of the susceptibilities is the microscopic content of
the Ginzburg--Landau description of Ref.~\cite{Banerjee2025}, in which the time-reversal-odd
component of the $d$--$f$ mixing is favored over the time-reversal-even one by a
phenomenological ratio of couplings $\lambda$, and the ordered phase requires $\lambda>1$.
Here the condition is not assumed but follows from the band structure, and the eigenvalue
condition $U_{\rm df}\chi_{\tau_2}=1$ derived below is its quantitative form. In the language of the $d$--$f$ pseudospin
$\tfrac12\Psi\dg\hat{\boldsymbol\tau}\Psi$, whose $x$ and $y$ components are the real and
imaginary parts of $\langle d\dg f\rangle$ and whose $z$ component is the $d$--$f$ occupation
difference, the symmetric phase already has a finite $x$-component, and the transition tilts
the pseudospin out of the $xz$-plane. Because the rotation about $z$ is not a symmetry, only
the sign of the $y$-component is free, and the order parameter is therefore an Ising one. Importantly, this circumstance allows us to avoid the dimensional restrictions enforced by the Mermin-Wagner theorem \cite{Mermin1966,Hohenberg1967}. 

In Section~II we have mentioned that $\Delta_{PT}$ and the surface magnetization $M_z$ belong to one
symmetry channel. The linearized theory must therefore be formulated on that channel rather
than on $\Delta_{PT}$ alone, and this requires the coupling that acts on the spin of the
$f$-electrons. 
It is the exchange interaction of the $f$-shell, $I_{\rm ff}$, the Hund's-rule
coupling of Ref. \cite{Roy2014} between
the two Kramers doublets of the $\Gamma_8$ quartet,
\beg
H_{\rm H}=-2J_{\rm H}\sum_{\mathbf R}\hat{\mathbf S}_{\mathbf R1}\cdot\hat{\mathbf S}_{\mathbf R2},
\en
where $\hat{\mathbf S}_{\mathbf R\alpha}=\sum_{\sigma\sigma'}f\dg_{\mathbf R\alpha\sigma}
\boldsymbol{\hat\sigma}_{\sigma\sigma'}f_{\mathbf R\alpha\sigma'}/2$. This interaction is not removed by the projection onto the $U_{\rm ff}\to\infty$ limit, since
it acts between different $\Gamma_8$ orbitals. Its longitudinal part can be written as
$-2J_{\rm H}\hat S^z_1\hat S^z_2=-({J_{\rm H}}/2)\big[\hat{\cal O}_M^2-\hat m_1^2-\hat m_2^2\big]$
with $\hat m_\alpha=2\hat S^z_\alpha$, where the last two terms reduce to a shift
of $\epsilon_f$ and to the intra-doublet repulsion already contained in the
mean field approximation. The uniform $f$-spin channel is therefore governed by
$-{I_{\rm ff}}\hat{\cal O}_M^2$ with $I_{\rm ff}=J_{\rm H}/4$ at the bare
level. Importantly, $I_{\rm ff}$ is a
separate coupling and is not fixed by $U_{\rm df}$. On each layer we thus retain two operators,
\begin{align}
\hat{\cal O}_\Delta(j)&=\ttau{2}\otimes\hs_0\otimes\hat{\mathbbm 1}_2,\nonumber\\
\hat{\cal O}_M(j)&=\tfrac12\big(1-\ttau{3}\big)\otimes\hs_z\otimes\hat{\mathbbm 1}_2,
\label{eq:m-ops}
\end{align}
both summed over the four orbital--spin components of the site, both are described by irreducible representation $A_2$ under $C_{4v}$ and
odd under time reversal: $\langle\hat{\cal O}_\Delta\rangle$ is the order parameter of
Eq.~\eqref{eq:DPT} and $\langle\hat{\cal O}_M\rangle$ represents the $f$-spin polarization per site.
The interaction projected on these two channels takes the form
\beg
H_{\rm int}=-\frac{U_{\rm df}}{2}\sum_j\hat{\cal O}_\Delta^2(j)-\frac{I_{\rm ff}}{2}\sum_j\hat{\cal O}_M^2(j).
\label{eq:m-Hint}
\en
The mean-field Hamiltonian is 
\beg\label{LayerMF}
\begin{aligned}
H_{\rm mf}(\bk_\parallel)&=H_{\rm slab}(\bk_\parallel)\\&+\sum\limits_{j=0}^{N-1}[\Delta_j\hat{\cal O}_\Delta(j)
+h_j\hat{\cal O}_M(j)]\otimes|j\rangle\langle j|. 
\end{aligned}
\en
In this expression $\Delta_j$ is the layer profile of $\Delta_{PT}$ and $h_j$ the layer profile of the exchange field
acting on the $f$-spins. Both of these quantities are determined from the following the self-consistency conditions
\beg\label{Self-Consistency}
\begin{aligned}
\Delta_j&=-U_{\rm df}T\sum\limits_{i\omega_n}\int\frac{d^2\bk_{\parallel}}{(2\pi)^2}\,{\rm Tr}\left[
\hat{\cal O}_\Delta \hat{G}_{S}(i\omega_n)\right]_{jj},\\
h_j&=-{I_{\rm ff}T}\sum\limits_{i\omega_n}\int\frac{d^2\bk_{\parallel}}{(2\pi)^2}\,{\rm Tr}\left[
\hat{\cal O}_M \hat{G}_{S}(i\omega_n)\right]_{jj},
\end{aligned}
\en
where $i\omega_n=i\pi T(2n+1)+\veps_F$ and $[\hat{G}_S]_{jj'}$ has to be computed from Eq. \eqref{GS} with $H_{\rm slab}$ replaced with $H_{\rm mf}$, Eq. \eqref{LayerMF}. The other time-reversal-odd on-site channels the surface admits, the
$A_1$ combination $\ttau{1}\otimes\hs_z$ and the in-plane $E$ doublet, are decoupled from
$\{\hat{\cal O}_\Delta,\hat{\cal O}_M\}$ by symmetry and can be examined separately with the
same machinery as they do not mix with the $A_2$ sector at the linear level.

\begin{figure}
\includegraphics[width=\linewidth]{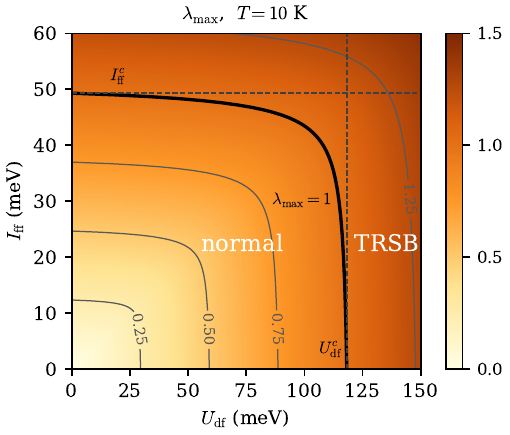}
\caption{Largest eigenvalue $\lambda_{\max}$ of the linearized self-consistency
kernel $\hat U^{1/2}\hat\chi\hat U^{1/2}$, Eq. \eqref{LinearSystem},  with $\hat U={\rm diag}(U_{\rm df},I_{\rm ff})$
and $\hat\chi_{ab}(j,j')$ the static layer susceptibility in the
$\{\hat O_\Delta,\hat O_M\}$ channel, as a function of the two couplings $U_{\rm df}$ and $I_{\rm ff}$, Eq. \eqref{eq:m-Hint}, at
$T=10$~K. The thick black line $\lambda_{\max}=1$ separates the normal
(time-reversal-symmetric) surface from the time-reversal-symmetry-broken (TRSB)
one. Thin grey lines are iso-contours of $\lambda_{\max}$, and the dashed lines
mark the single-channel thresholds $U_{\rm df}^{c}=118$~meV (at $I_{\rm ff}=0$)
and $I_{\rm ff}^{c}=49$~meV (at $U_{\rm df}=0$). Model parameters used to generate this plot are: $t_d=-2$, $t_f=0.05$, $V=0.0365$~eV, $\eta_d=-0.3$,
$\eta_f=-0.29$, $\varepsilon_d=4.4$~eV, $\varepsilon_f=-0.1165$~eV,
$E_F=-6.35$~meV. We have also included the band bending potential $U(j)=U_0e^{-ja/\lambda_B}$ with $U_0=-1.5$~meV,
$\lambda_B=2a$ ($a$ is the interlayer distance) and $N=18$ explicit layers coupled to the semi-infinite bulk through
$\hat\Sigma$. The momentum integrals have been evaluared using the $96\times96$ points $\mathbf k_\parallel$ mesh over the surface Brillouin
zone. Lastly, the Matsubara cutoff $\Lambda=11$~eV with the analytic high-frequency tail has also been used.}
\label{Fig-lam-contour}
\end{figure}

\section{Results}\label{SectionIV}
The calculations that we present below proceed in three steps, each of which uses the output of the previous one.
We start by solving the linearized self-consistency problem, Eqs.~\eqref{LinearSystem} and
\eqref{Susceptibility}, in the time-reversal invariant state. The requirement that the leading eigenvalue
of the layer-resolved susceptibility kernel reaches unity at $T_c\approx10$~K fixes the
relation between the couplings $U_{\rm df}$ and $I_{\rm ff}$, and the corresponding
eigenvector gives the layer structure of the incipient order and shows whether it is bound
to the surface. Having determined these couplings we then we proceed with solving the full self-consistency equations
\eqref{Self-Consistency} below $T_c$ and obtain the profiles $\Delta_j$ and $h_j$ as
functions of temperature. Lastly, we use the resulting apex propagator to compute the
tunneling current, Eq.~\eqref{eq:IV}, and its magnetic contrast, Eq.~\eqref{Eq4Integrand},
which we then compare with the experimental data.
\subsection{Linear susceptibility}
The values of the  interaction parameters $U_{\rm df}$ and $I_{\rm ff}$ will have to be fixed before we can compute the order parameters at a given temperature which then can be used in the calculation of the tunneling current. 
First we expand the right hand side of \eqref{Self-Consistency} to the first order in powers of $\Delta_j$ and $h_j$ yields the following system of linear equations
\beg\label{LinearSystem}
\begin{aligned}
\Psi_{aj}=\eta_a\sum\limits_{bj'}\chi_{aj,bj'}(T)\Psi_{bj'}
\end{aligned}
\en
where we introduced the shorthand notations $\vec{\Psi}_{j}=(\Delta_j,h_j)$, $\vec{\eta}=(U_{\rm df}, I_{\rm ff})$ and the susceptibility matrix
\beg\label{Susceptibility}
\begin{aligned}
\chi_{aj,bj'}(T)&=-T\sum\limits_{i\omega_n}\int\frac{d^2\bk_{\parallel}}{(2\pi)^2}\\&\times{\rm Tr}\left\{
\hat{\cal O}_a [\hat{G}_{S}(i\omega_n)]_{jj'}\hat{\cal O}_b [\hat{G}_{S}(i\omega_n)]_{j'j}\right\}.
\end{aligned}
\en
In \eqref{Susceptibility} matrix function $\hat{G}_{S}$ is determined by \eqref{GS}. Equations \eqref{LinearSystem} define a linear system. Since experimentally the time-reversal symmetry breaking happens at $T\approx 10$\,K, by fixing temperature to this value and evaluating the dependence of the linear system's eigenvalues on $U_{\rm df}$ and $I_{\rm ff}$. For a given value of $I_{\rm ff}$ by fixing temperature and requiring that after symmetrizing $\chi_{aj,bj'}$ the maximum eigenvalue equals unity we find the corresponding value of $U_{\rm df}$. In Fig. \ref{Fig-lam-contour} we present a contour plot of a maximum eigenvalue $\lambda_{\rm max}$ of \eqref{Susceptibility} for different values of the interaction parameters. It is important to emphasize here that both channels cannot condense separately, i.e. nonzero value of $\Delta_j$ necessarily drives the condensation of $h_j$ and vice versa. This confirms the general form of Eq. \eqref{eq:mj} above. 

\begin{figure}
\includegraphics[width=\linewidth]{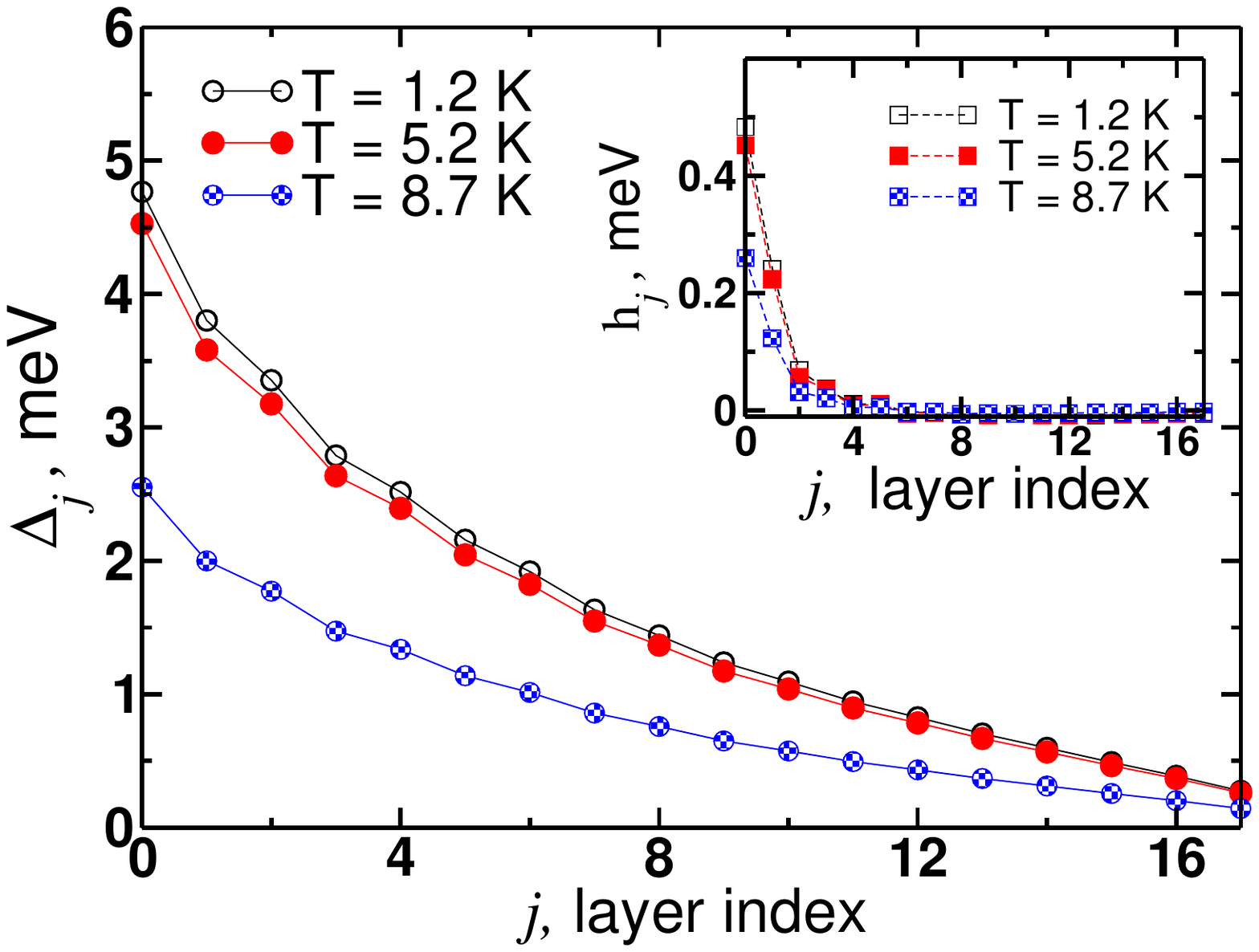}
\caption{Self-consistent layer profiles of the two components of the surface
order parameter at $U_{\rm df}=116.8$~meV, $I_{\rm ff}=20$~meV (a point on the
$T_c=10$~K line of Fig.~\ref{Fig-lam-contour}): the odd-parity hybridization
amplitude $\Delta_j$ (main panel) and the $f$-shell Weiss exchange field $h_j$
(inset) for $T=1.2$, $5.2$ and $8.7$~K, $j=0$ being the apex layer. The
profiles are obtained by iterating Eqs.~(\ref{Self-Consistency}) to convergence on
$N=18$ explicit layers coupled to the nonmagnetic semi-infinite bulk, with the
same model parameters, band-bending potential and $\mathbf k_\parallel$ mesh
as in Fig.~\ref{Fig-lam-contour} above. Notably, the
exchange field $h_j$ is confined to the first three to four layers, whereas
$\Delta_j$ decays over the whole explicit region, reflecting the
predominantly bulk-derived character of the hybridization channel.}
\label{Fig-DLTj-hj}
\end{figure}

\subsection{Temperature dependence of the order parameter components}
Having mapped the region of couplings in which the surface layers of the tip
break time-reversal symmetry, we fix $U_{\rm df}=116.8$~meV and
$I_{\rm ff}=20$~meV, a point on the $\lambda_{\max}=1$ line of
Fig.~\ref{Fig-lam-contour} at which the instability sets in at $T_c
\simeq 9.8$~K. The
choice of a finite $I_{\rm ff}$, well below its own threshold, is motivated by
the small static surface magnetization inferred from the
experiment~\cite{Banerjee2025}; the hybridization channel then drives the
transition and the exchange field follows. With these couplings we solve the
self-consistency equations~\eqref{Self-Consistency} for the two components
$\Delta_j$ and $h_j$ of the order parameter in the layers between the apex
and the time-reversal-invariant bulk. We show their dependence on the layer index 
in Fig.~\ref{Fig-DLTj-hj}. We find that the exchange field $h_j$ is confined to the first
three to four layers, where the band bending is strongest, whereas
$\Delta_j$ decays slowly over the whole intermediate region. The two length
scales express the feature that we have already found in the linear analysis: the surface pins
the instability, but the hybridization channel is predominantly bulk-derived. Indeed, as we have checked, the bulk crystal has its own TRSB instability which 
requires a critical value of the interaction $U_{\rm df}^c\approx 120$ meV while the critical value for the instability which is confined to the surface requires 
a critical value $U_{\rm df}\approx 118$ meV: it is only a few percent below the bulk threshold value $U_{\rm df}^c$.

\begin{figure}
\includegraphics[width=0.95\linewidth]{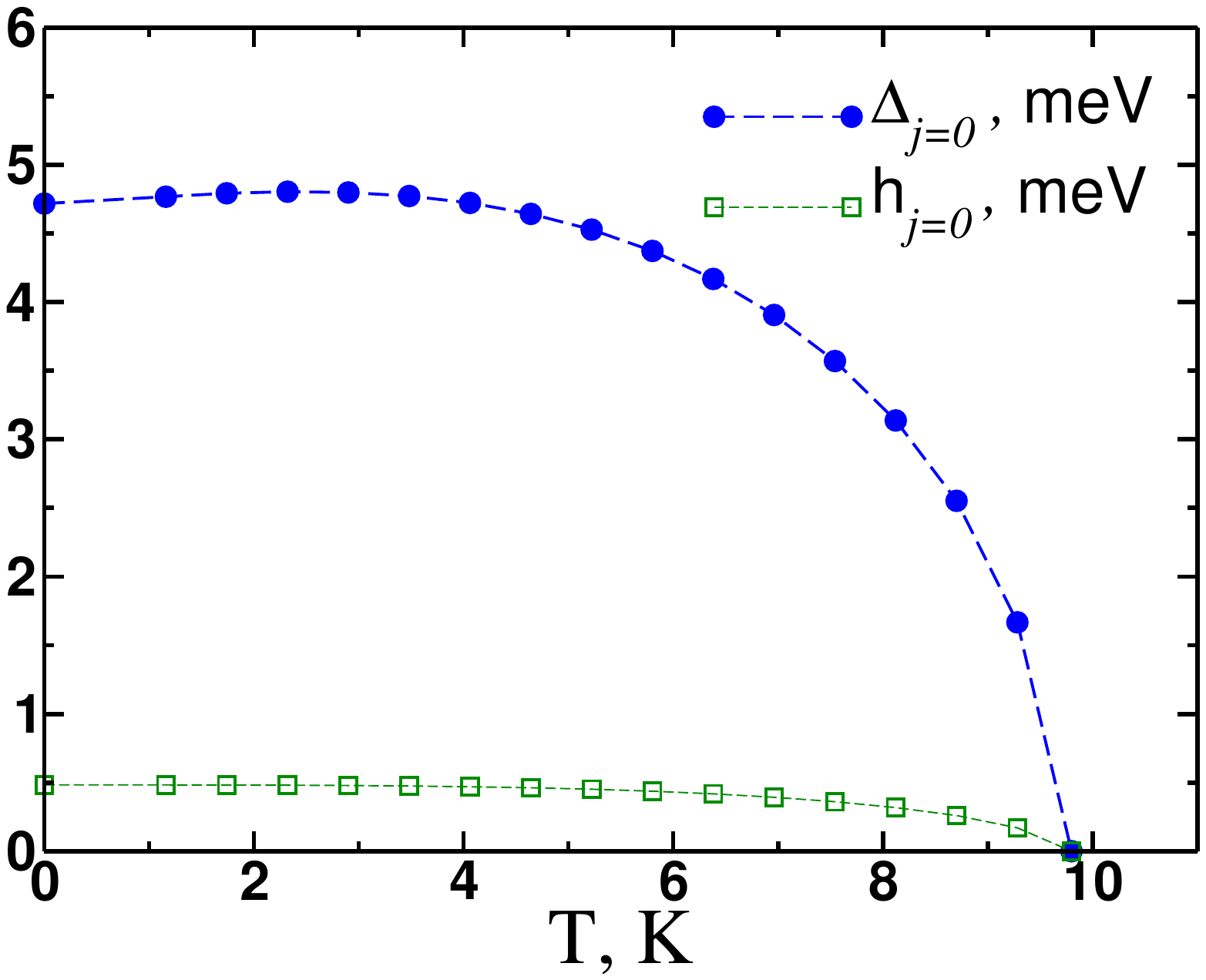}
\caption{Temperature dependence of the order parameters $\Delta_{j=0}$ and exchange field $h_{j=0}$ on the apex extracted from the solution of the self-consistency equations
\eqref{Self-Consistency}. The parameters used to generate this plot are the same as those used for Fig. \ref{Fig-DLTj-hj}.}
\label{Fig-DLT0-h0}
\end{figure}

Finally, in Fig. \ref{Fig-DLT0-h0} we show the temperature dependence of the order parameters $\Delta_0$ and $h_0$ at the apex layer $j=0$. Both vanish continuously at $T_c\simeq 9.8$~K with
the mean-field exponent, $\Delta_0,h_0\propto\sqrt{T_c-T}$, and saturate
below $T\approx 3$~K. Notably, the ratio
$h_0/\Delta_0\simeq 0.10$ is the same at every temperature and equals the
ratio of the two components in the eigenvector of the linearized problem:
the exchange field and the odd-parity hybridization are locked together as
one order parameter in agreement with the arguments based on the symmetry analysis above. The saturated amplitude is large on the scale of the
transition temperature, $\Delta_{j=0}(T=0)/k_BT_c\simeq 5.7$, far above the
weak-coupling value $1.76$. We interpret this as a direct consequence of the bulk-derived
character of the hybridization channel. Finally, the exchange field $h_0\simeq 0.5$~meV corresponds to an
$f$-shell spin polarization $\langle\hat O_M(0)\rangle=-h_0/I_{\rm ff}\simeq
-0.02$ per apex site, i.e.\ a static moment of $\sim 0.03\mu_B$, which is consistent with the bound on the bias-independent contrast extracted
from the experimental data \cite{Banerjee2025}.

\subsection{Tunneling conductance}
The self-consistent fields $\Delta_j$ and $h_j$ obtained in the previous
subsection are all we need to compute the tunneling spectra. As we have
shown in Section~II (see also Appendix~\ref{AppendixA}), the current and its
magnetic contrast depend on the tip only through the Hermitian matrix
$\hat{\cal A}_\br(\omega)$. For a point-like apex the tunneling matrix
$\hat T_\br$ projects on the apex site, and the calculation of
$\hat{\cal A}_\br$ reduces to that of the spectral matrix of the outermost
layer, $j=0$, weighted by the tunneling amplitudes $t_d$ and $t_f$ of the
$d$- and $f$-orbitals~\cite{Maltseva2009}:
\beg\label{A00Weighted}
\big[\hat{\cal A}_\br(\omega)\big]_{\sigma\sigma'}=\sum_{a=d,f}t_a^2
\sum_{\alpha\beta}\big[\tilde A_{00}(\omega)\big]_{a\alpha\sigma,\,a\beta\sigma'},
\en
where $\alpha$ and $\beta$ label the two orbitals of the $d$- or
$f$-quartet, both of which we assume to couple to the substrate with the
same amplitude, and
\beg\label{A00GRA}
\tilde{A}_{00}(\omega)=\frac{i}{2\pi N_\parallel}\sum_{\bk_\parallel}
\left[\hat{G}_S^R(\bk_\parallel,\omega)-\hat{G}_S^A(\bk_\parallel,\omega)\right]_{00}
\en
is the spectral matrix of the apex layer averaged over the surface
Brillouin zone, $N_\parallel$ being the number of points of the
$\bk_\parallel$ mesh. Equation~\eqref{A00Weighted} contains no terms
proportional to $t_dt_f$, which would describe the interference between
tunneling into the $d$- and into the $f$-states \cite{Maltseva2009}. These terms vanish
identically in both phases. The product of time reversal and a reflection
in a vertical mirror plane, which is a symmetry of the pristine surface, is
also preserved by the $A_2$ order parameter. This symmetry makes the
on-site $d$--$f$ block of $\tilde A_{00}$ purely imaginary, so that its
contribution to the current cancels for any real $t_d$ and $t_f$.

\begin{figure}
\includegraphics[width=\linewidth]{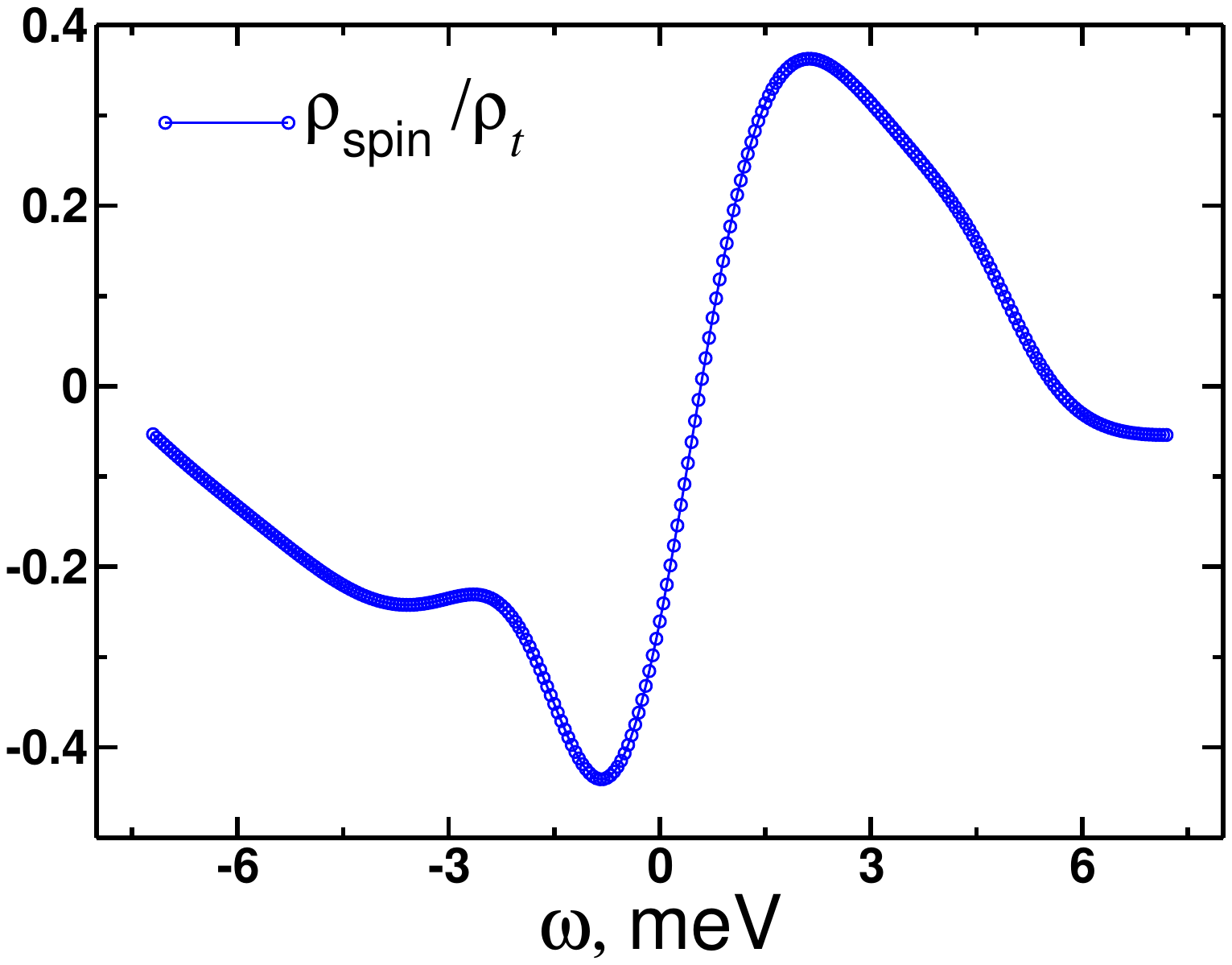}
\caption{Spin polarization of the apex, $\rho_{\rm spin}/\rho_t$, at
$T=5.2$~K, computed with the self-consistent fields $\Delta_j$ and $h_j$ at
this temperature and with equal tunneling amplitudes, $t_d=t_f$. Both
$\rho_{\rm spin}$ and $\rho_t$ are convolved with $-\partial f/\partial\omega$
at the same temperature. The spectra are obtained with $\gamma=0.15$~meV on
the same $N=18$ layers and $\bk_\parallel$ mesh as in
Fig.~\ref{Fig-DLTj-hj}.}
\label{Fig-rhospin-rhot}
\end{figure}

In what follows we insert the solution of Eqs.~\eqref{Self-Consistency}
into the layer Hamiltonian \eqref{LayerMF} and evaluate the retarded
function $\hat G_S^R$ on the real frequency axis, replacing $\omega$ by
$\omega+i\gamma$ with $\gamma=0.15$~meV. The parameter $\gamma$ accounts in
a phenomenological way for the finite lifetime of the quasiparticles, which
in a real tip comes from processes not included in our mean-field theory,
such as inelastic electron--electron scattering and scattering on
imperfections of the apex. Owing to the fourfold rotational symmetry of the
surface the matrix $\hat{\cal A}_\br$ is diagonal in spin, and we denote
its two diagonal elements, the densities of states of the apex for the two
spin projections on the surface normal, by $\rho_\up$ and $\rho_\dn$. Two
quantities enter the tunneling current. The first is the total density of
states of the apex,
\beg\label{rhot}
\rho_t(\omega)=\rho_\up(\omega)+\rho_\dn(\omega)=\mathrm{Tr}_\sigma\hat{\cal A}_\br(\omega),
\en
which determines the part of the current that does not change when the tip
is moved by half the magnetic period,
$\delta I_{\br,\br+\bR_{{1}/{2}}}^{\rm even}$, and hence the ordinary
conductance. The second is the spin-resolved density of states,
\beg\label{rhospin}
\rho_{\rm spin}(\omega)=\rho_\up(\omega)-\rho_\dn(\omega)
=\mathrm{Tr}_\sigma\big[\hs_z\hat{\cal A}_\br(\omega)\big],
\en
which determines the magnetic contrast
$\delta I_{\br,\br+\bR_{{1}/{2}}}^{\rm odd}$ through
Eq.~\eqref{eq:nogo-contrast}, where according to Eq.~\eqref{Eq4Integrand}
it appears multiplied by $\hat{\mathbf n}\cdot\hat{\mathbf z}$, the
projection of the substrate moments on the normal to the tip surface.

In Figure~\ref{Fig-rhospin-rhot} we show the ratio $\rho_{\rm spin}/\rho_t$ as a
function of bias at $T=5.2$~K, where both $\rho_{\rm spin}$ and $\rho_t$
have been convolved at this temperature using
\beg\label{Convolve}
\rho(eV)=\int\limits_{-\infty}^\infty \rho(\omega)\left[-\frac{\partial f(\omega)}{\partial \omega}\right]_{\omega-eV}d\omega
\en
before taking the ratio. If the density of states and the polarization of
the substrate do not vary on the scale of a few meV, the ratio of the
magnetic to the ordinary part of the differential conductance at bias
$eV=\omega$ is equal to this quantity multiplied by the bias-independent
factor $(\rho_1/\rho_0)(\hat{\mathbf n}\cdot\hat{\mathbf z})
\cos(\mathbf Q\cdot\mathbf r+\varphi_0)$ of Eq.~\eqref{Eq4Integrand}. Below
$\omega\simeq0.6$~meV the polarization is negative and reaches its largest
magnitude, $0.43$, at $\omega\simeq-0.85$~meV. It is positive between
$0.6$ and $5.6$~meV, with a maximum of $0.36$ at $\omega\simeq2.1$~meV, and
becomes negative again at higher bias. At zero bias the polarization is
finite, $\rho_{\rm spin}/\rho_t\simeq-0.26$, whereas in the absence of
time-reversal symmetry breaking $\rho_{\rm spin}$ vanishes identically.
These results are not sensitive to the ratio of the tunneling amplitudes:
varying $t_f/t_d$ between $0.1$ and $1$ alters $\rho_{\rm spin}/\rho_t$ by
no more than $0.05$ and shifts the first sign change by less than
$0.02$~meV.

\begin{figure}
\includegraphics[width=\linewidth]{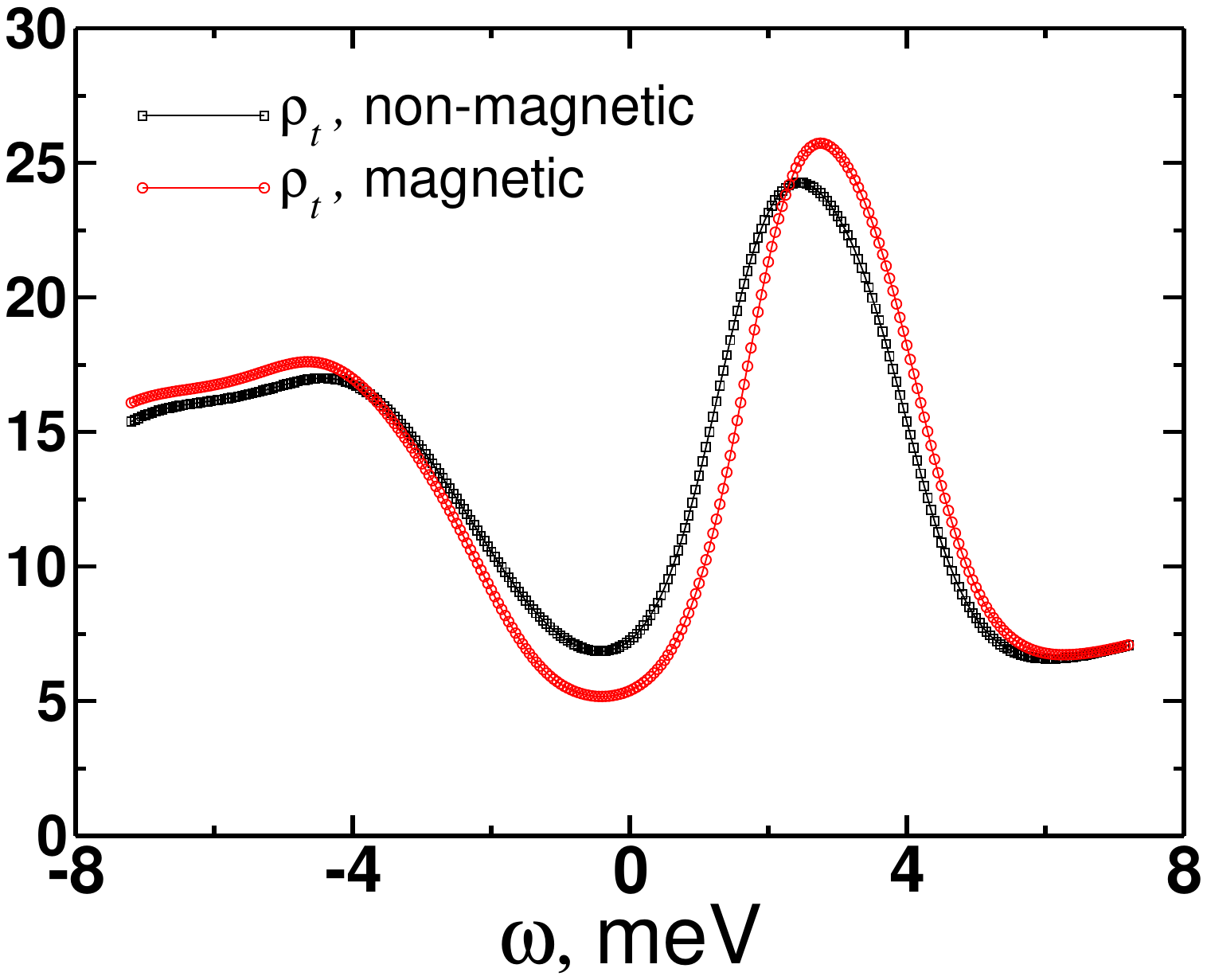}
\caption{Total density of states of the apex, $\rho_t$, in units of
eV$^{-1}$ per apex site for $t_d=t_f=1$, for the non-magnetic
apex ($\Delta_j=h_j=0$, black squares) and for the magnetic state at
$T=5.2$~K (red circles). All other
parameters are the same as in Fig.~\ref{Fig-rhospin-rhot}.}
\label{Fig-rhot}
\end{figure}

The origin of this structure becomes clear once we look at the surface
Dirac cones in the magnetic state. The time-reversal-odd fields lift the
Kramers degeneracy at the time-reversal-invariant momenta of the surface
Brillouin zone. At $\bar\Gamma$ the lower of the two levels, at
$-0.57$~meV, has spin down and the upper one, at $1.72$~meV, has spin up,
whereas at $\bar X$ and $\bar Y$ the order is reversed, with spin up at
$-2.23$~meV and spin down at $-0.77$~meV. The cones thus acquire masses of
$1.14$~meV at $\bar\Gamma$ and $0.73$~meV at $\bar X$ and $\bar Y$, and
these masses have opposite signs. Between $-0.77$ and $-0.57$~meV the upper
bands of the $\bar X$ and $\bar Y$ cones and the lower band of the
$\bar\Gamma$ cone carry the same spin, and the polarization is largest in
magnitude close to this interval. Since the Fermi level lies inside the gap
of the $\bar\Gamma$ cone, the spin-resolved density of states there comes
entirely from the vicinity of $\bar X$ and $\bar Y$. The positive maximum
near $2$~meV, on the other hand, comes from the vicinity of $\bar\Gamma$,
where the upper level of the $\bar\Gamma$ cone merges with the edge of the
bulk conduction band. At higher bias the contribution of $\bar X$ and
$\bar Y$ again exceeds that of $\bar\Gamma$, and the polarization changes
sign for the second time. In contrast to the case of a single cone,
Eq.~\eqref{eq:oddstep}, the sign changes of $\rho_{\rm spin}$ are therefore
decided by the competition between cones whose masses have opposite signs,
and their positions are not pinned to the energy of any Dirac point. As a
consequence, the bias at which the polarization first changes sign depends
on temperature already through the thermal broadening. Keeping the fields
fixed at their values for $T=5.2$~K, we find that it moves from $0.8$~meV
for the broadening corresponding to $T=3.5$~K to $0.2$~meV for that
corresponding to $T=8.7$~K.

In Figure~\ref{Fig-rhot} we compare the total density of states \eqref{rhot} of
the magnetic and of the non-magnetic apex, both convolved with
$(-\partial f/\partial\omega)$ at $T=5.2$~K and are therefore proportional to
the ordinary differential conductance at bias $eV=\omega$. In the magnetic state $\rho_t$
is reduced between $\omega\simeq-3.7$ and $2.3$~meV, an interval close to
the bulk gap, $-3.0<\omega<2.5$~meV, in which both Dirac points lie, and it
is enhanced outside this interval. The reduction is $26\%$ at zero bias and
reaches its largest value, $30\%$, at $\omega\simeq0.8$~meV, inside the gap
opened at the $\bar\Gamma$ Dirac point, while the peak at the edge of the
bulk conduction band moves from $2.45$ to $2.75$~meV.

\begin{figure}
\includegraphics[width=\linewidth]{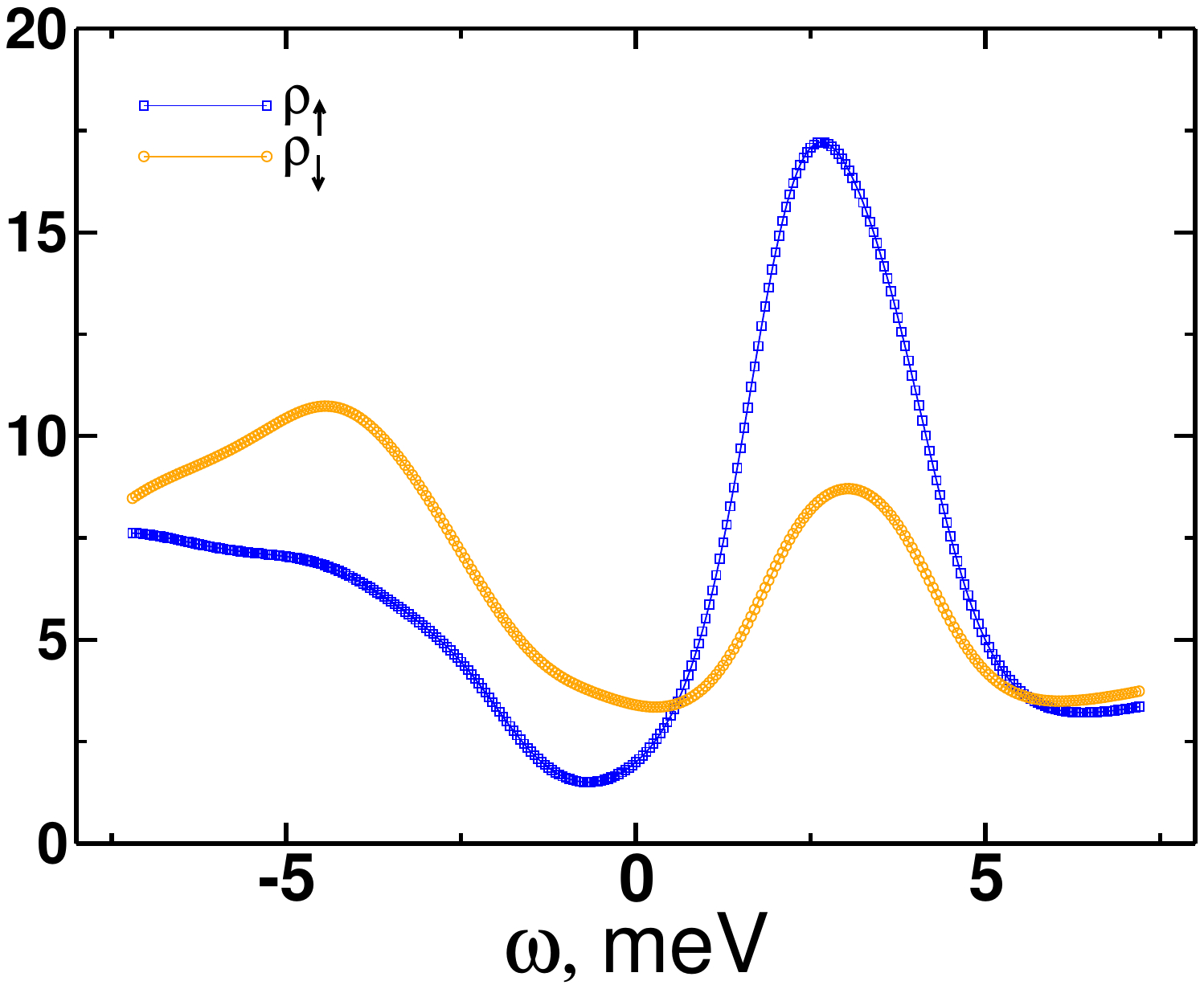}
\caption{Densities of states of the apex for the two spin projections on the
surface normal, $\rho_\up$ (blue squares) and $\rho_\dn$ (orange circles), in
the magnetic state at $T=5.2$~K, in units of eV$^{-1}$ per apex site for
$t_d=t_f=1$. Both are convolved using \eqref{Convolve} at this
temperature. In the non-magnetic apex $\rho_\up=\rho_\dn=\rho_t/2$, with $\rho_t$
shown in Fig.~\ref{Fig-rhot}. All other parameters are the same as in
Fig.~\ref{Fig-rhospin-rhot}.}
\label{Fig-rhoup-rhodn}
\end{figure} 

Finally, in Figure \ref{Fig-rhoup-rhodn} we show the bias dependence of the spin-resolved density of states at $T=5.2$ K. 
Since $\rho_t\pm\rho_{\rm spin}=2\rho_{\up,\dn}$, for a substrate fully
polarized along the normal to the tip surface the conductance measured above
the two magnetic sublattices of the substrate would be proportional to
$\rho_\up$ and $\rho_\dn$, so that the figure shows the largest contrast the
ordered state can produce. Within the window of Fig.~\ref{Fig-rhoup-rhodn}
each spin channel retains its spectral weight to within a few percent, and
the order mainly moves the weight in energy, spin-down states accumulating
below the Fermi level and spin-up states above it. The two largest features
come from the $\bar\Gamma$ cone, whose lower band, with spin down, merges
with the edge of the bulk valence band and whose upper band, with spin up,
merges with the edge of the conduction band. Compared with the pristine apex,
$\rho_\dn$ is enhanced by $26\%$ at $\omega\simeq-4.5$~meV and $\rho_\up$ by
about $45\%$ between $2.6$ and $4$~meV. The absolute contrast is therefore
largest near the edge of the conduction band, where at
$\omega\simeq2.5$~meV $\rho_\up$ exceeds $\rho_\dn$ by a factor of two. Close
to the Fermi level, inside the gap of the $\bar\Gamma$ cone, the spin-down
states of the $\bar X$ and $\bar Y$ cones dominate, and the reduction of
$\rho_t$ at zero bias found above comes almost entirely, to $88\%$, from the
spin-up channel. Finally, the $f$-spin density of the occupied states within
$10$~meV of the Fermi level, $-0.017$ per apex site, accounts for about three
quarters of the polarization
$\langle\hat{\cal O}_M(0)\rangle=-h_0/I_{\rm ff}\simeq-0.023$ obtained in
Sec.~IV.B, so that the static moment is carried mostly by the states close
to the Fermi level.

\section{Discussion}
% Discuss possibility of nonlinear contributions; also this \ref{Hlawenka2018}
We would like to emphasize that our theory rests on the bulk hybridization between the
$d$- and $f$-orbitals. Specifically, the channel that becomes unstable at the surface is the
time-reversal-odd component of the same $d$--$f$ orbital overlap that forms the bulk
gap, and the band bending potential at the surface merely pins the instability to the first few
layers. This bears on the debate over the nature of the surface states of
SmB$_6$~\cite{Dzero2010,Dzero2012,DzeroGalitski2013,Dzero2016,Kim2014,Syers2015,Tan2015,Hlawenka2018,Eo2019}. A single helical Dirac cone with a Kramers
degenerate node admits exactly one on-site mass term, and that term is odd
under time reversal. A topologically protected surface can therefore gap its states only
by breaking time-reversal symmetry, which is precisely what the tunneling
data show it does, at the temperature at which the surface states lose their
spectral weight~\cite{Aishwarya2022}. Trivial surface states, in contrast,
can be gapped by time-reversal-even terms and would have no reason to order.
While the bias reversal does not by itself prove the topological character
of the surface states, it is what a topological surface must do and what a
trivial one has no reason to do.

This statement has a compact form in the language of axion electrodynamics,
which Banerjee \emph{et al.}~\cite{Banerjee2025} used to interpret the
experiment. The bulk of a strong topological insulator has the axion angle
$\vartheta=\pi$ and the vacuum has $\vartheta=0$, so that the surface of the
tip is a domain wall across which $\vartheta$ changes by $\pi$. In the Dirac
description $\vartheta$ is the phase of a complex mass whose real part is the
time-reversal-even mass of the band inversion and whose imaginary part is the
parity- and time-reversal-odd mass $\Delta_{PT}$ of Eq.~\eqref{eq:DPT}. When
$\Delta_{PT}=0$ this mass stays real and passes through zero at the surface, so
that the helical surface states are bound to this zero as in an Ising domain
wall. When $\Delta_{PT}\neq 0$ the mass rotates in the complex plane from
$\vartheta=\pi$ to $\vartheta=0$ without vanishing, as the magnetization does
in a Bloch wall, and the surface states become massive. Banerjee \emph{et al.}
attributed this rotation to the imaginary part of the on-site $d$--$f$
hybridization. Our theory provides a microscopic realization of that idea.
Specifically, we did not assume the Bloch wall from the start but found
that it forms spontaneously below $T_c$, as shown in Fig.~\ref{Fig-DLTj-hj}.
Importantly, in our theory this domain wall has two parts. In the outer three
to four layers, where the band bending is strongest, $\Delta_{PT}$ is
accompanied by the exchange field $h_j$ and the magnetization $M_z$, as the
$A_2$ symmetry of the surface requires. Deeper inside, the environment of each
layer is nearly cubic. Under the cubic group $O_h$ the operator of
Eq.~\eqref{eq:DPT} transforms according to $A_{1u}$ and $M_z$ is a component
of the axial vector $T_{1g}$, and the two fall into the same representation
$A_2$ only when the symmetry is lowered to $C_{4v}$. Their coupling is
therefore a surface effect, which is why $h_j$ decays much faster than
$\Delta_j$. In our solution the Fermi level lies inside the gap of the
$\bar\Gamma$ cone but above the gaps of the $\bar X$ and $\bar Y$ cones, so
that the ordered surface is a metal with broken time-reversal symmetry and its
anomalous Hall and magnetoelectric responses are not quantized~\cite{Elio2016}.

The axionic interpretation of Ref. ~\cite{Banerjee2025} and ours differ also in the origin of the bias
reversal, and this difference can be tested experimentally. In
Ref.~\cite{Banerjee2025} the spin polarization of the apex is induced by the
electric field in the junction through the magnetoelectric response. The bias
$V$ drops almost entirely across the vacuum gap between the tip and the substrate,
so that the corresponding electric field acting on the apex is of order $V/d$, where $d$ is the width of the gap, and
at fixed bias it decreases as the tip is retracted. The contrast then changes
sign at the bias $V^*$ at which the induced moment cancels any static one. In our theory the moment is static, and
the sign change is a property of the equilibrium spectrum of the apex,
$\rho_{\rm spin}(\omega)$, which the conductance samples at $\omega=eV$. However, in our theory magneto-electric 
contribution is also allowed by symmetry, because the $A_2$ order parameter also
permits a linear magnetoelectric response, and the question is which of them
controls the sign change. We believe that this question can be answered experimentally by measuring $V^*$ at several
set-point currents. In our theory the magnetic and the total conductance are
both proportional to the tunneling probability, which drops out of the ratio
$\rho_{\rm spin}/\rho_t$, so that neither the polarization nor $V^*$ should depend
on the position of the tip. If the moment is induced by the
field, on the other hand, the induced part grows as the tip approaches the
substrate at fixed bias, so that $V^*$ is either pinned at zero bias, when
there is no static moment, or moves with the set-point. 

The same distinction bears on how the data of Ref.~\cite{Banerjee2025} are analyzed. There the spin polarization of the tip is written as $m_t(\omega,V)=m_0+VX(\omega)$, with $m_0$ a bias-independent constant given by the zero-bias intercept, $m_0=I'(0)/e$, of the contrast current $I(V)$. In our theory, however, the static polarization of the apex is not a constant but the function $\rho_{\rm spin}(\omega)$ shown in Fig.~\ref{Fig-rhospin-rhot}, and the intercept is governed by its value at the Fermi energy rather than by a separate static moment. Let us write to linear order in the field $\rho_{\rm spin}(\omega;V)=\rho_{\rm st}(\omega)+VX(\omega)$, insert it into Eq.~\eqref{eq:nogo-contrast} with $\rho_1$ smooth and $T=0$, so that $I(V)\propto\int_0^{eV}\rho_{\rm st}(\omega;V)\,d\omega$, and divide both sides by $eV$. If the static part $\rho_{\rm st}(\omega)$ were a constant, we would immediately find $X(eV)=d[I(V)/eV]/dV$, which is the inversion formula used in Ref.~\cite{Banerjee2025}. For the actual spectrum $\rho_{\rm st}(\omega)$ the same operation gives
\begin{equation}
\begin{aligned}
X_{\rm ext}(eV)&=X(eV)+\frac{\rho_{\rm st}(eV)}{V}-\frac{1}{eV^2}\int\limits_0^{eV}\rho_{\rm st}(\omega)\,d\omega .
\end{aligned}
\label{eq:Xext}
\end{equation}
We thus see that only a constant static spectrum is removed and the energy dependence of $\rho_{\rm st}(\omega)$ reappears as an apparent field-induced response. Near zero bias the last two terms in Eq.~\eqref{eq:Xext} together tend to $(e/2)\rho_{\rm st}'(0)$, the intercept is $m_0=\rho_{\rm st}(0)$, and the contrast changes sign at $V^*\simeq-\rho_{\rm st}(0)/[e\rho_{\rm st}'(0)+2X(0)]$, so that
\begin{equation}
\frac{m_0}{X_{\rm ext}(0)}=-2V^*,
\label{eq:ratio}
\end{equation}
regardless of how the spin-resolved spectral function of the tip is split between the static and the field-induced parts. Equation~\eqref{eq:ratio} indicates that the ratio of the intercept to the extracted response near zero bias measures the position of the sign change and not the magnitude of the magneto-electric response. We therefore conclude that a bias spectrum taken at a single tip position cannot by itself separate the static and the field-induced contributions to the magnetic contrast. In particular, replacing the constant $m_0$ by an energy-dependent static term does not remove this difficulty: for any field-induced $X(\omega)$ there is then a static term that produces the same contrast at every bias, so that $X(\omega)$, and with it the magnitude of a magnetoelectric response, is not determined by such spectra. It is the dependence on the set-point current discussed above that separates them, and we leave a detailed treatment of this problem for future work.

Our results also show that the transition can actually be detected without a
magnetic substrate. In the point-contact geometry of Section~IV.C the
total density of states of the apex $\rho_t$, Eq. \eqref{rhot}, enters the tunneling
current through its product with the density of states of the substrate,
$\rho_0$, and this part of the current does not require the substrate to be
magnetic. An SmB$_6$ tip brought to the surface of a normal metal whose
density of states is featureless on the scale of a few meV therefore
measures $\rho_t$ directly, and Fig.~\ref{Fig-rhot} shows that the in-gap
conductance of such a junction must decrease when the surface orders. The
size of this decrease, however, cannot be used by itself to identify the transition.
At $T=5.2$~K the zero-bias conductance of the magnetic apex is $26\%$ below
that of the nonmagnetic apex at the same temperature, but the nonmagnetic apex is
not accessible below $T_c$. The thermal broadening, which fills the minimum
of $\rho_t$ near the Fermi level, also decreases on cooling, and for fixed
band parameters it alone lowers the zero-bias conductance of the nonmagnetic
apex by about $30\%$ between $T_c$ and $5.2$~K. The two effects can
nevertheless be separated, because $\rho_t$ is the same for the two domains
$\pm\Delta_{PT}$ and is therefore an even function of the order parameter.
Close to $T_c$ its change is proportional to $\Delta_{PT}^2\propto T_c-T$,
whereas the thermal broadening changes smoothly with temperature, so that
the zero-bias conductance as a function of temperature must show a change of
slope at $T_c$. In contrast with the nonmagnetic case, the magnetic contrast, which is odd in $\Delta_{PT}$, increases as $(T_c-T)^{1/2}$. However, the validity of both statements is limited by the constraints of the mean-field theory.
Including the fluctuations effects of the Ising order beyond mean-field will likely change the singularity in the conductance
into that of the internal energy, as for any quantity which is even in the
order parameter, but they do not fully remove it. For this reason we expect that in realistic experimental situations the reduction of the in-gap conductance may be smaller.

\section{Conclusions}
% Provide brief summary: (a) no helical tunneling; (b) time reversal breaking; (c) time reversal breaking leads to bias reversal 
In this work we have shown by explicit microscopic calculation that the
$d$--$f$ hybridization, whose onset in the bulk is a smooth crossover,
acquires at the surface a time-reversal-odd component $\Delta_{PT}$ which
appears through a genuine phase transition. Being a spin scalar,
$\Delta_{PT}$ transforms according to a one-dimensional representation of
the surface point group, so that it is an Ising order parameter and the
only magnetization it can induce is the component $M_z$ along the surface
normal, which belongs to the same representation. At the apex and at the
lowest temperatures we find $\Delta_{PT}\simeq 5$~meV and a Weiss field
$h_z\simeq 0.5$~meV, and the ratio of the two does not change with
temperature.

The conclusion that time-reversal symmetry is broken on the surface of
SmB$_6$ does not rest on this model. A spin-polarized tunneling signal
requires a nonzero spin-resolved spectral function of the apex, and no
time-reversal-invariant Hamiltonian of the tip---whatever the spin--orbit
coupling, the band bending or the structure of its surface states---can
produce such a function, so that the magnetic contrast observed in
Refs.~\cite{Aishwarya2022,Banerjee2025}, together with its bias reversal,
is by itself evidence of a broken symmetry. The model enters only in the
answer to the question of how the symmetry is broken: the surface removes
inversion, the hybridization is allowed to become complex, and its
imaginary part, driven by the inter-orbital repulsion $U_{\rm df}$ and assisted
by the exchange $I_{\rm ff}$, condenses below $T_c$. 
For the couplings chosen here this happens at $T_c\simeq 9.8$~K, in agreement with the temperature above which the magnetic contrast disappears in the experiment. The bulk, whose own threshold lies a few percent higher,
remains time-reversal symmetric at all temperatures, and even if it ordered
the same operator would belong in the cubic bulk to a representation which
carries no magnetic moment, so that the magnetism we describe is a property
of the surface alone.

The ordered state has a definite signature in the tunneling spectra. The
order parameter lifts the Kramers degeneracy of the surface Dirac points
and gives the cones at $\bar\Gamma$ and at $\bar X$, $\bar Y$ masses of
opposite signs, about $1.1$ and $0.7$~meV at $T=5.2$~K. The competition
between these cones produces a spin polarization of the apex of up to about
$40\%$, which changes sign within about $1$~meV of zero bias. That spin polarization also changes sign at a bias
that moves with temperature and remains finite at zero bias. At the same
time the ordinary conductance inside the bulk gap is reduced by about a
quarter, an effect which does not require a magnetic substrate.

The cornerstone of our theory is that the TRSB transition at the surface is driven by the opening of the bulk hybridization gap. Therefore the surface states
which carry the magnetic signal must be those which form together with the
gap: a trivial surface band detached from the hybridization would be
unaffected by the transition and would show no contrast reversal at $T_c$.
The order parameter $\Delta_{PT}$ is odd under both inversion and time
reversal, so that the surface of SmB$_6$ below 10~K realizes an odd-parity
magnet in which the primary order is not a magnetization but a
time-reversal-odd component of the hybridization, $M_z$ appearing only as
its by-product. The Ising character of the order implies domains of
opposite $M_z$ on the same facet, and whether the same mechanism operates
on other facets, and in other Kondo insulators whose surfaces break
inversion, are questions which the present calculation makes concrete and
which we leave for future work.

\begin{acknowledgments}
We thank Piers Coleman and Vidya Madhavan for useful discussions. We are also grateful to Piers Coleman for sharing the revised version of Ref.~\cite{Banerjee2025} with us before it became publicly available. The work of MD was financially supported by the National Science
Foundation Grant No. DMR-2400484. MD thanks the Niels Bohr Institute,
where a part of this work was completed, for its hospitality. 
The authors acknowledge the use of Anthropic's
Claude for cross-checking the analytical derivations, for independent
numerical verification of the results, and for figure preparation and
manuscript checking. All outputs of Claude were verified and validated by
the authors.
\end{acknowledgments}

\section*{Data availability} The data that support the findings of this article were generated by numerical evaluation of the analytical expressions derived herein and can be regenerated in full from the information given in the paper. For this reason the authors did not  deposit them in a public repository. The data and the codes that produce figures presented in this manuscript are available from the authors upon request. 

\begin{appendix}
\begin{widetext}
%%%% APPENDIX A
\section{Tunneling current}\label{AppendixA}
The general aspects of the tunneling theory are well documented \cite{Bardeen1961,Caroli1971,Julliere1975,Tersoff1985,Meservey1994,Mahan,Chen2008}. We
summarize them here mainly for the purpose of fixing the conventions.
We start by writing the total Hamiltonian as a sum of three terms: 
\beg\label{eq:Eq1a}
H = H_{\rm tip} + H_{\rm s} + H_T, 
\en
where $H_{\rm tip}$ described the tip, $H_{\rm s}$ describes the substrate,
and $H_T$ accounts for the coupling across the vacuum barrier.
In what follows we will peep the spin indices explicit rather than summing them away at the start. The tunneling part of the Hamiltonian is:
\begin{equation}
H_T \;=\; \sum_{\bk,\bp}\sum_{\alpha\beta}
\Big[\,T_{\beta\alpha}(\mathbf p,\bk)\;\hat{s}\dg_{\mathbf p\beta}\,\hat{t}_{\bk\alpha}\;+\;\text{h.c.}\Big].
\label{eq:HT}
\end{equation}
Here $\hat{t}(\hat{s})$ are fermionic annihilation operators for the tip (substrate), the values of momentum $\bk$ run over the tip's surface Brillouin zone, the
values of momentum $\mathbf p$ run over the substrate's surface Brillouin zone, while the indices $\alpha$ and $\beta$ each run over the two values $\{\uparrow,\downarrow\}$. Therefore $T(\mathbf p,\bk)$ is a $2\times2$ matrix in spin space, with row index $\beta$ on the
substrate side and column index $\alpha$ on the tip side. Nothing else in the problem carries spin, so \eqref{eq:HT} is where all of the spin physics of
the junction resides.

In order to derive the expression for the tunneling current we will assume that tunneling matrix element is small and develop the perturbative expansion to the second order in $H_T$, which corresponds to the first order in the tunneling probability. It is worth noting that this approach has also been adopted for this problem by other groups \cite{Aishwarya2022,Banerjee2025}.
A single tunnelling event takes one electron from a tip single-particle eigenstate $|\nu\rangle$
of energy $\varepsilon_\nu$ into a substrate single-particle eigenstate $|\lambda\rangle$ of
energy $\xi_\lambda$. The index $\nu$ runs over all tip single-particle eigenstates, $\nu = 1,\dots,D_t$, with spin
already included in that count. Consequently, the index $\lambda$ runs over all substrate single-particle eigenstates,
$\lambda = 1,\dots,D_s$. Introducing the single-particle matrix element $M_{\lambda\nu} = \langle\lambda|H_T|\nu\rangle$ with averaging running over the ground state of the tip and substrate, we use the Fermi's Golden Rule to obtain the expression for the tunneling rate from tip to substrate:
\begin{equation}
\Gamma_{t\to s} \;=\; \frac{2\pi}{\hbar}\sum_{\nu=1}^{D_t}\sum_{\lambda=1}^{D_s}
|M_{\lambda\nu}|^2\;f_t(\varepsilon_\nu)\,\big[1-f_s(\xi_\lambda)\big]\;
\delta(\xi_\lambda-\varepsilon_\nu),
\end{equation}
where 
\begin{equation}
M_{\lambda\nu} \;=\; \sum_{\bk{\mathbf p}}\sum_{\alpha\beta}
\langle\lambda|\mathbf p\beta\rangle\;T_{\beta\alpha}(\mathbf p,\bk)\;
\langle\bk\alpha|\nu\rangle .
\end{equation}
and the reverse rate $\Gamma_{s\to t}$ by exchanging $t\leftrightarrow s$ in the occupation
factors. In the net current the blocking factors cancel, since
$f_t(1-f_s)-f_s(1-f_t)=f_t-f_s$, so for the net tunneling current we find
\begin{equation}
I \;=\; \frac{2\pi e}{\hbar}\sum_{\nu=1}^{D_t}\sum_{\lambda=1}^{D_s}
|M_{\lambda\nu}|^2\,\big[f_t(\varepsilon_\nu)-f_s(\xi_\lambda)\big]\,
\delta(\xi_\lambda-\varepsilon_\nu).
\label{eq:Igr}
\end{equation}
Taking into account that external bias has been applied for the subsequent calculations it is convenient to re-write \eqref{eq:Igr} as follows
\begin{equation}
I \;=\; \frac{2\pi e}{\hbar}\int d\omega\,\big[f_t(\omega+eV)-f_s(\omega)\big]
\sum_{\nu=1}^{D_t}\sum_{\lambda=1}^{D_s}
|M_{\lambda\nu}|^2\,\delta(\omega+eV-\varepsilon_\nu)\,\delta(\omega-\xi_\lambda),
\label{eq:Iomega}
\end{equation}
which is the form we will use in what follows. 

Next we introduce two spectral functions for the two respective leads:
\begin{align}
\big[A_t(\omega)\big]_{\bk\alpha,\,\bk'\alpha'} &=
\sum_{\nu=1}^{D_t}\langle\bk\alpha|\nu\rangle\langle\nu|\bk'\alpha'\rangle\,
\delta(\omega-\varepsilon_\nu),\\
\big[A_s(\omega)\big]_{\mathbf p'\beta',\,\mathbf p\beta} &=
\sum_{\lambda=1}^{D_s}\langle\mathbf p'\beta'|\lambda\rangle\langle\lambda|\mathbf p\beta\rangle\,
\delta(\omega-\xi_\lambda).
\end{align}
Using $T^*_{\beta'\alpha'}(\mathbf p',\bk')=[T\dg]_{\alpha'\beta'}(\bk',\mathbf p')$ we then compactly write the sum in \eqref{eq:Iomega} as
\begin{equation}
{\;
\sum_{\nu=1}^{D_t}\sum_{\lambda=1}^{D_s}|M_{\lambda\nu}|^2\,
\delta(\omega-\varepsilon_\nu)\,\delta(\omega-\xi_\lambda)
\;=\;\Tr\!\big[\,T\,A_t(\omega)\,T\dg\,A_s(\omega)\,\big],\;}
\label{eq:tracekey}
\end{equation}
with the trace over the composite index momentum $\otimes$ spin.
Equation \eqref{eq:tracekey} is the golden-rule sum over the spin states at both ends which has been written in a matrix form.
Since each lead is translationally invariant along its own surface, so $A_t$ is diagonal in $\bk$ and
$A_s$ is diagonal in $\mathbf p$. Then the composite trace collapses to
\begin{equation}
\sum_{\bk\mathbf p}\Tr_\sigma\!\big[\,T(\mathbf p,\bk)\,A_t(\bk,\omega)\,
T\dg(\mathbf p,\bk)\,A_s(\mathbf p,\omega)\,\big],
\end{equation}
where $\Tr_\sigma$ is the $2\times2$ spin trace, that is
$\sum_{\alpha,\beta\in\{\uparrow,\downarrow\}}$ and the expression for the tunneling current reads:
\begin{equation}
I(V) \;=\; \frac{2\pi e}{\hbar}\sum_{\bk{\mathbf p}}\int d\omega\,
\big[f(\omega+eV)-f(\omega)\big]\;\Tr_\sigma\!\big[\,T(\mathbf p,\bk)\,A_t(\bk,\omega+eV)\,T\dg(\mathbf p,\bk)\,
A_s(\mathbf p,\omega)\,\big].
\label{eq:IV}
\end{equation}
Equation \eqref{eq:nogo-current} in the main text is based on this expression for the current.

%%%% APPENDIX B 
\section{Spin-resolved density of states of a massive Dirac cone}\label{AppendixB}
Let us consider one surface Dirac cone with the Hamiltonian
\beg
\hat{h}_j(\bq)=\eps_j+\mathbf d_j(\bq)\cdot\hat{\mbox{\boldmath $\sigma$}},
\label{eq:B-h}
\en
where $\eps_j$ is a node energy, $\mathbf d_j(\bq)=\big(v_jq_y,\,-v_jq_x,\,m_j\big)$, $v_j$ is velocity, $m_j$
is mass due to time-reversal symmetry breaking and $\hat{\mbox{\boldmath $\sigma$}}$ are Pauli matrices. Since $({\mathbf d}_j\cdot\hat{\mbox{\boldmath $\sigma$}})^2=E_{j}^2(\bq)$, the eigenvalues of \eqref{eq:B-h} are $\eps_j\pm E_j({\bq})$. Furthermore, the
projectors onto the two bands are
\beg
\hat \Pi_\pm=\tfrac12\big(1\pm\hat{\mathbf d}_j\cdot\hat{\mbox{\boldmath $\sigma$}}\big),\qquad
\hat{\mathbf d}_j=\frac{{\mathbf d}_j}{E_j},
\label{eq:B-P}
\en
as one verifies from $\hat \Pi_\pm^2=\hat \Pi_\pm$ and $({\mathbf d}_j\cdot\hs)\hat \Pi_\pm=\pm E_{j}\hat \Pi_\pm$.
The spin expectation value in either band is therefore
\beg
\Tr\big[\hat{\mbox{\boldmath $\sigma$}}\,\hat \Pi_\pm(\bq)\big]
=\pm\frac{1}{E_j({\bq})}\big(v_jq_y\hat{\mathbf x}-v_jq_x\hat{\mathbf y}+m_j\hat{\mathbf z}\big).
\label{eq:B-spin}
\en
The in-plane components of the trace are the helical texture, odd in $\bq$, and the component along
the mass axis is $\pm m_j/E_j$, even in $\bq$, opposite in the two bands.

The spectral matrix of the Dirac cone at energy $\omega$ is
\beg
\hat \Sigma_j(\bq,\omega)=\hat \Pi_+(\bq)\,\delta[\omega-\eps_j-E_j({\bq})]
+\hat \Pi_-(\bq)\,\delta[\omega-\eps_j+E_j({\bq})],
\label{eq:B-A}
\en
Taking the trace with $\hs_z$ and using Eq.~\eqref{eq:B-spin} for the spin part of the spectral function it obtains
\beg
\rho_{\rm spin}^{(j)}(\omega)
=\int\!\frac{d^2q}{(2\pi)^2}\,\frac{m_j}{E_{\bq}}\Big[\delta(\omega-\eps_j-E_{\bq})-\delta(\omega-\eps_j+E_{\bq})\Big].
\label{eq:B-mspin}
\en
We note that the helical components drop out of the trace with $\hs_z$ and they also drop out of the
$\bq$-integral, being odd in $\bq$. The integrand depends on $\bq$ only through $E_{\bq}$,
so with $d^2q=2\pi q\,dq$ and $E\,dE=v_j^2\,q\,dq$ we have
\beg
\int\!\frac{d^2\bq}{(2\pi)^2}\,F(E_{\bq})=\frac{1}{2\pi v_j^2}\int\limits_{|m_j|}^{\infty}\!E\,F(E)dE,
\label{eq:B-measure}
\en
which is the statement that the density of states of the cone is $E/2\pi v_j^2$ per unit
area outside the gap. The factor $1/E$ of the spin projection cancels the factor $E$ of the
density of states, and the delta functions fix $E=|\omega-\eps_j|$:
\beg
\rho_{\rm spin}^{(j)}(\omega)=\frac{m_j}{2\pi v_j^2}\Big[\theta(\omega-\eps_j-|m_j|)-\theta(\eps_j-\omega-|m_j|)\Big]
=\frac{m_j}{2\pi v_j^2}\,{\rm sgn}(\omega-\eps_j)\,\theta\big(|\omega-\eps_j|-|m_j|\big),
\label{eq:B-result}
\en
which is Eq.~\eqref{eq:oddstep} given in the main text. The total density of states, obtained from
Eq.~\eqref{eq:B-A} with $\hs_0$ in place of $\hs_z$, is $|\omega-\eps_j|/2\pi v_j^2$ outside
the gap and is even about the node. In contrast, the spin-resolved one is odd, and its magnitude is
constant because the polarization of a state, $m_j/E$, falls off exactly as the number of
states grows.

Three comments are in order: (i) the result above holds for either sign of $m_j$ and either helicity of the
cone, so reversing $m_j$ reverses $\rho_{\rm spin}$; (ii) with a finite lifetime broadening $\gamma$ the step edges at
$|\omega-\eps_j|=|m_j|$ are smeared over $\gamma$ and the value inside the gap becomes of
order $\gamma m_j/(2\pi v_j^2|m_j|)$ rather than zero, but the sign change at $\omega=\eps_j$
is exact, since $\hat A_j(\eps_j+\delta)$ and $\hat A_j(\eps_j-\delta)$ are related by the
particle--hole map $\hs_y\hat A_j^{*}\hs_y$ that reverses $\hs_z$; (iii) what the substrate
reads is $\Tr[(\hat{\mathbf n}\cdot\hs)\hat A_j]=(\hat{\mathbf n}\cdot\hat z)\,\rho_{\rm spin}$,
since the in-plane components integrate to zero. As a result the apex tunneling weight of the cone and
the orbital vertex multiply the whole expression and are supplied by the microscopic
calculation of Section \ref{SectionIV}.
\end{widetext}
\end{appendix}

\bibliography{references}

\end{document}